\documentclass{jfm}

\usepackage{graphicx}
\usepackage{newtxtext}
\usepackage{newtxmath}
\usepackage{natbib}

\usepackage{graphicx}
\usepackage{dcolumn}
\usepackage{bm}
\usepackage[bbgreekl]{mathbbol}
\usepackage{overpic}
\usepackage{dcolumn}
\usepackage{epstopdf}
\usepackage{amsmath}  
\usepackage{amssymb}  
\usepackage{bm}  
\usepackage{siunitx}
\usepackage{mathrsfs}
\makeatletter
\def\amsbb{\use@mathgroup \M@U \symAMSb}
\makeatother
\expandafter\let\csname equation*\endcsname\relax
\expandafter\let\csname endequation*\endcsname\relax
\usepackage{amsmath}
\usepackage{subcaption,caption}

\newcommand{\ve}[1]{\boldsymbol{#1}}
\newcommand{\ma}[1]{\ensuremath{\mathbb{#1}}}

\newcommand{\tr}{{\rm Tr}}

\usepackage{hyperref}
\hypersetup{
    colorlinks = true,
    urlcolor   = blue,
    citecolor  = black,
}

\newcommand{\RomanNumeralCaps}[1]

\title{Bifurcations in Stokes Flow Sedimentation}

\author{Elias Huseby\aff{1,2},
  Pierre Mathier\aff{3},
 Meera Das\aff{3},
 Arjun Menezes\aff{3},
 Theo Witkamp\aff{4},
  Ziqi Wang\aff{4},
 Bernhard Mehlig\aff{5},
 \and Greg A. Voth\aff{3}\corresp{\email{gvoth@wesleyan.edu}}
 }

\affiliation{\aff{1} Environmental Fluid Mechanics Laboratory, Stanford University, Palo Alto, CA 94305, USA  
\aff{2} Computational and Mathematical Engineering, Stanford University, Palo Alto, CA 94305, USA
\aff{3}Department of Physics, Wesleyan University, Middletown, CT 06459, USA

\aff{4}Department of Applied Physics and Science Education,
Eindhoven University of Technology, 5600 MB Eindhoven, Netherlands
\aff{5}Department of Physics, Gothenburg University, 41296 Gothenburg, Sweden
}

\begin{document}
\maketitle

\begin{abstract}

Particles whose shapes couple translation to rotation display a rich array of behaviors as they sediment at low Reynolds number. We introduce a unifying perspective in which the possible dynamical regimes and bifurcations between them can be understood. We use experimental measurements of helical ribbons, with controlled center of mass offsets, to identify the key bifurcation from complex dynamics to a single attracting state as the magnitude of the offset increases.   
The sedimentation dynamics are very sensitive to small center of mass offsets, with the bifurcation occurring for offsets less than one percent of the particle length.   Using mobility tensors obtained from immersed boundary method simulations, we simulate helical particle sedimentation and identify 
 an alignment bifurcation surface, defined in the three dimensional space of center of mass offsets, that separates simple from complex sedimentation dynamics. Inside this surface we find limit cycles which emerge through Hopf and homoclinic bifurcations. Cocentered particles with coincident centers of force and mobility provide a reference case at the center of the bifurcation surface.  We show how the geometric and dynamical symmetries of sedimenting cocentered particles are broken as the center of force offset moves away from the cocentered case.
Three parity time-reversal (PT) symmetries exist for all cocentered particles under reflections normal to the eigenvectors of its translation-rotation coupling tensor.  When a center of force offset preserves at least one of these PT symmetries, then there are closed orbits for particles inside the alignment bifurcation surface.   

\end{abstract}




\section{Introduction}

Consider a particle of arbitrary shape falling through a still fluid at low Reynolds number. What trajectory does it trace as it falls?   Simple homogeneous particles in the shape of spheres or ellipsoids are constrained by their symmetry to have no coupling between translation and rotation. As a consequence, their orientation remains fixed as they settle, and they nearly instantly reach a constant velocity. It is precisely because of this simplicity that spheres and ellipsoids are the standard first approximations when modeling the transport of particles in environmental, industrial, and biological systems~\citep{Bala2010,DiBenedetto2026}. 
When a particle shape couples translation to rotation, spatial trajectories during sedimentation can be quite complex~\citep{miara_dynamics_2024,huseby2025helical,joshi2025sedimentation}, and it can be difficult to connect geometry to dynamics. 

\begin{figure}
    \centering
    \includegraphics[trim = {0 2.5cm 0 0},clip,width=0.75\linewidth]{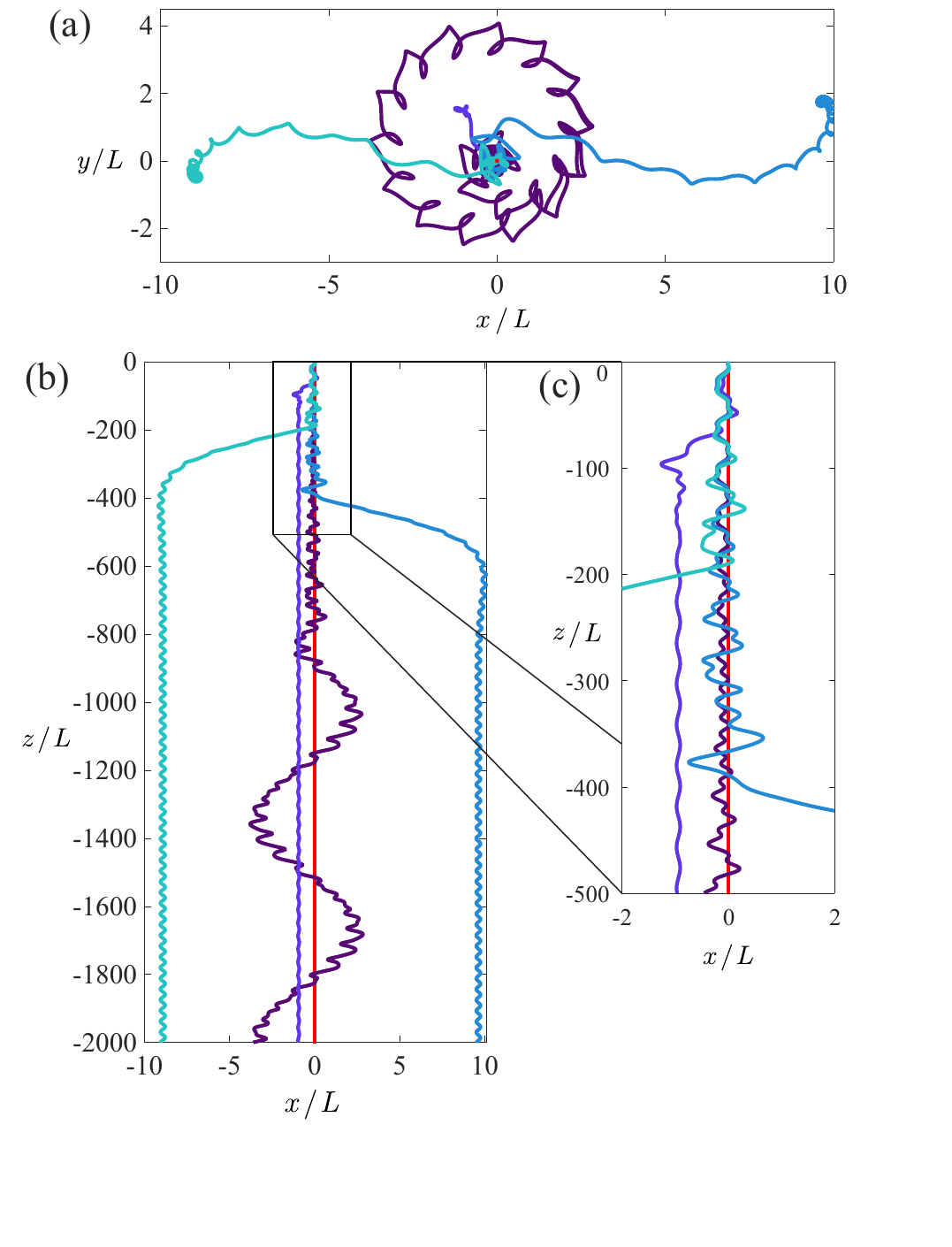}
    \caption{Spatial trajectories of five helical ribbons with identical shape and initial orientation. Panels show (a) top and (b--c) side views, with trajectories distinguished by color. The red trajectory is that of an idealized cocentered particle (of length $L$), while the rest have a center of mass offset of $0.002L$ in 4 randomly chosen directions. All particles are initialized with $\hat{\mathbf y}$ aligned with gravity, corresponding to Euler angles $(\theta,\psi)=(0,\pi/2)$.}
    \label{fig:shape_perturbed}
\end{figure}

To illustrate the wide variety of dynamics that can occur for nearly identical particles, we show in Figure~\ref{fig:shape_perturbed} the sedimentation dynamics of a helical ribbon, along with four other particles with identical shape whose center of mass has been offset by only 0.2\% of the particle length in four randomly chosen directions. This offset is of the same order of magnitude as the offset error between the actual and planned center of mass of particles when they are 3D printed on a FormLabs Form 2 printer~\citep{huseby2025helical}. It can be seen in Figure~\ref{fig:shape_perturbed} that, for some particles, their long-time behavior emerges nearly immediately; for others, it develops only after falling nearly a thousand body lengths. The dramatically different dynamics introduced by these small perturbations shows the importance of understanding how the settling dynamics of typical, imperfect, particles relate to the settling behavior of highly symmetric particles that have been widely studied. 

Consideration of geometric symmetries of particles and dynamic symmetries of the equations of motion have been central to most efforts to comprehend translation-rotation coupling.  Reflection symmetry breaking has been a primary consideration since its importance to the rotation of light polarization as it translates past molecules was identified~\citep{pasteur1848asymmetry,barron2009molecular}.  The word chirality, which means no reflection symmetries~\citep{kelvin1894molecular}, has become deeply connected to translation-rotation coupling since light is not rotated by solutions of non-chiral molecules when they are randomly oriented. However, focus on chirality has often led people astray. The polarization of light can be rotated even by non-chiral molecules if they are held in fixed orientations~\citep{hobden1967} and translation-rotation coupling generally does not require reflection symmetry breaking~\cite{Efrati}.  Breaking of reflection symmetry is neither necessary nor sufficient for translation-rotation coupling.  Instead, reflection symmetries provide strong constraints on the types of translation-rotation coupling that can occur. Since the word chiral does not tell us anything about whether or not a particle has translation-rotation coupling, we need new terminology. Here we use the term {\em helical} for particles with non-zero translation-rotation coupling due to their shape and {\em non-helical} for particles for which this coupling is absent.

Separately, research has often focused on the role of symmetry of the equations of motion under combined reflection and time reversal symmetry, called parity-time reversal (PT) symmetry~\citep{Bender1998,Elganainy2018}. Low Reynolds number hydrodynamics provides a concrete setting where the role of PT symmetry and PT symmetry breaking can be understood more clearly.

The dynamics of sedimenting particles are best visualized in a two-dimensional space of orientations that neglects rotations about the gravity vector.    Different choices of these two variables are used in the literature, including Euler angles, Tait-Bryan angles, and the trajectory of the gravitational vector on the unit sphere in body coordinates.   \citet{gonzalez_dynamics_2004} identify the fixed points of the two dimensional dynamics as the key feature for understanding sedimentation dynamics and they survey many cases. For a typical particle that does not have any special symmetries, they identify two possibilities: either (1) there are two fixed points with one attracting and one repelling and particles converge to a single orientation in the 2D space (while possibly also rotating around the gravity vector), or (2) there are six fixed points with two attracting, two saddles, and two repelling and the long dynamics depends sensitively on the initial orientation.   Most of the particles that have been studied have special geometric symmetries and show a wider range of long term dynamics than these two typical cases.

\begin{figure*}
    \centering
    \includegraphics[width=1\linewidth]{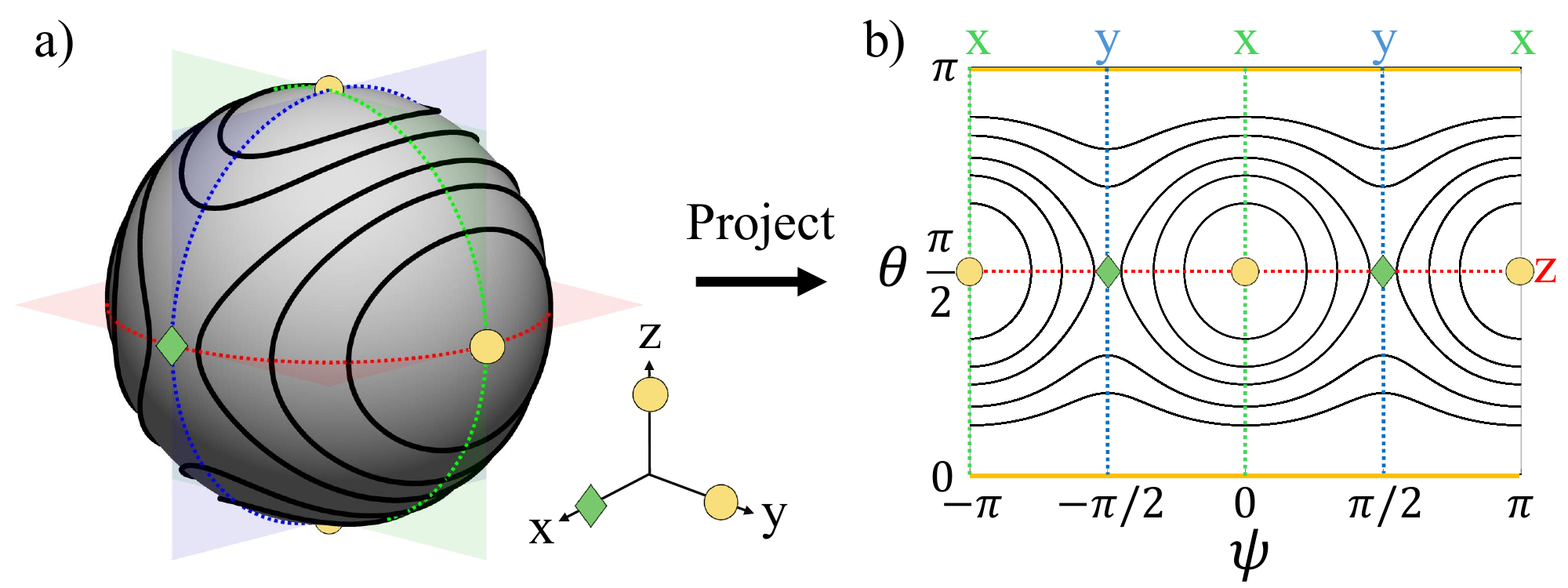  }
    \caption{Orientation dynamics of a cocentered particle.   (a) gravity vector trajectories in the particle's body coordinate system (b) Euler angles of tilt ($\theta$) and spin ($\psi$).  Yellow circles are centers.  Green diamonds are saddles.   Planes of parity-time reversal symmetry are labeled according to their normal vector in body coordinates, which is the center of mass offset direction under which the symmetry is preserved: (green) $\hat{x}$ plane, (blue) $\hat{y}$ plane, (red) $\hat{z}$ plane. In the phase plane, yellow bars at $\theta=0$ and $\theta=\pi$ correspond to the centers along $\hat{z}$ which are stretched into lines by the mapping. }
    \label{fig:cocentered}
\end{figure*}

In this paper, we offer a unifying perspective in which the dynamics of any sedimenting particle can be understood.  Any particle has a reference particle with the same surface geometry but with its center of force coincident with its center of mobility. This reference particle has simple dynamics shown in Figure~\ref{fig:cocentered} with an orientation that is either stationary or in closed orbits in orientation space due to three planes of PT symmetry. The actual particle can then be placed in the 3D parameter space given by the vector displacement between the center of force and the center of mobility in the eigenbasis of the reference particle's translation-rotation coupling tensor.

For small center of force offsets, we find six fixed points in experiments and simulations as expected~\citep{gonzalez_dynamics_2004,witten_review_2020}.  However, the dynamics are quite complex with a wide variety of nonlinear dynamics including limit cycles.  As the center of force moves farther from the center of mobility, a pair of saddle-node bifurcations occurs at which the dynamics become much simpler with one attracting and one repelling fixed point.   There is a surface in the space of center of force offsets at which the bifurcation occurs which we call the alignment bifurcation surface. We identify four different classes of cocentered reference particles which have quite different bifurcation surfaces. 

We use helical ribbons as simple particles with non-zero translation rotation coupling and whose center of force, mobility, and resistance are all the same when made of homogeneous material.   By moving the center of mass with metal beads epoxied into 3D printed plastic helical ribbons, we explore the space of center of force offsets.  In experiments, we move the center of force along each of the principal axes of the mobility tensors and observe the alignment bifurcation surface.  In simulations, we extend the exploration to the rest of the space and find many fascinating features of the dynamics, including limit cycles, Hopf and homoclinic bifurcations where the limit cycles appear, as well as co-dimension 2 lines where two different saddle-node bifurcations coincide.

\section{Theory}

\subsection{Mobility Formalism}

In Stokes flow, the disturbance velocity and pressure fields due to a particle are linear in the imposed driving. So a particle's velocity, angular velocity, rate of strain, force, torque, and stresslet are all linearly coupled through the particle's surface geometry ~\citep{Happel,kim2013microhydrodynamics}. 

Conventionally, problems are divided into those of resistance or mobility, with imposed motion or forces, respectively. For quiescent sedimentation where the force, $\ve f$, and torque, $\ve \tau$, are imposed and the strain rate is zero, a particle's velocity $\ve v$ and angular velocity $\ve \omega$ are given by:

\begin{equation}
\begin{bmatrix} \ve v \\ \ve \omega \end{bmatrix}  
= \frac{1}{\mu} 
\begin{bmatrix}\ma a' & {\ma b'}^{\top}  \\\ma b' & \ma c'  \end{bmatrix}
\begin{bmatrix}\ve f \\\ve \tau\end{bmatrix} 
\label{eqft}
\end{equation}
where the mobility tensors $\ma a$, $\ma b$, and $\ma c$ are primed to indicate they are in lab coordinates.

For sedimentation problems, there are two unique centers that are important: the center of mobility around which $\ma b$ is symmetric 
and the center of force around which $\ve \tau=0$~\citep{witten_review_2020}. 
The center of mobility is determined entirely by a particle's surface geometry, while for sedimentation in a uniform gravity field, the center of force is determined by the location of the center of mass, $\ve r_{cm}$, and the center of buoyancy, as well as the force applied at each center. 
The center of force, $\ve r$, in a coordinate system with the center of buoyancy at the origin is given by 
\begin{equation}
\begin{aligned}
    \ve r = \frac{\rho_p}{\rho_p -\rho_{f}}\ve r_{cm}  \qquad\qquad \ve f = V(\rho_p- \rho_{f})\ve g
\end{aligned}
\end{equation}
where $\rho_p$ is the particle density, $\rho_f$ is the fluid density, and $\ve g$ is the gravitational acceleration vector. For helical ribbons made of homogeneous materials, three rotational symmetries ensure that the center of mobility and the center of buoyancy are the same point.

The translation theorem
\begin{equation}
\ma b_{f} = \ma b_{m}+ \ma c \times\ve r
\label{b_trans}
\end{equation}
relates the translation-rotation coupling around the center of force ($\ma b_f$) and center of mobility ($\ma b_m$) separated by a displacement $\ve r$. The columns of $\ma c \times \ve r$ are the cross products between each column of $\ma c$ and the vector $\ve r$. A similar expression exists for $\ma a_f$~\citep{kim2013microhydrodynamics}.

Around the center of force, the particle's equations of motion are then 
\begin{subequations}
\label{eq:eom}
\begin{eqnarray}
\tfrac{{\rm d} }{{\rm d}t}{\ve x} &=& \ve v\,, \quad \mu \ve v = 
\ma R \; \ma a_f\; \ma R^{-1} \ve f \label{eq:vf} \\
\tfrac{{\rm d} }{{\rm d}t} \ma R&=& \ve \omega \times \ma R   \,,\quad \mu \ve \omega = 
\ma R \; {{\ma b_f}}  \; \ma R^{-1} \ve f \,.\label{eq:wf}
\end{eqnarray}
\end{subequations}

where \ma R is the rotation matrix from the body-fixed frame to the laboratory frame.
Note that the angular dynamics determine, but are independent from, the translation dynamics. So, even if spatial trajectories are of interest, the underlying angular dynamics should be the focus and these are determined by the properties of $\ma b_f$.

\subsection{Symmetries of the Orientation Dynamics}

To clarify the connection between particle geometry, mass distribution, and dynamics, it is useful to work in a body coordinate system defined by the orthogonal eigenvectors of $\ma b_m$. In body coordinates the normalized force vector $\hat{\ve g} = \ma R^{-1} \ve f/|\ve f|$ evolves as
\begin{equation}
    \frac{d}{dt}\hat{\ve g}=-\frac{|\ve f|}{\mu}(\ma b_f \hat{\ve g})\times \hat{\ve g}
    \label{eq:g}
\end{equation}
so that $\hat{\ve g}$ traces a path on the unit sphere as the particle rotates.  The reorientation timescale is set by $\mu/(|\ma b_f| \; |\ve f|)$,  the ratio of the viscosity to the product of the applied force and translation-rotation coupling strength. In this work, we primarily consider stability properties and symmetries; we therefore set $|\ve f|/\mu = 1$ unless otherwise indicated. This formulation naturally excludes rotations around $\ve f$ from the dynamics, so the fixed points of (\ref{eq:g}) correspond to steady rotations ($\frac{d}{dt}\ve \omega=0$) around each of the real eigenvectors of $\ma b_f$. 

The orientation dynamics of a particle are symmetric under a transformation $\ma Q $ if that transformation takes every solution curve to another solution of the same equations of motion (\ref{eq:g}). A time reversal symmetry requires the same condition, but with the dynamics backwards in time.

Mathematically, the condition is satisfied when
the equations of motion for a transformed state $\ma Q \hat{\ve g}$
\begin{equation}
    \frac{d}{dt}(\ma Q\hat{\ve g})=-(\ma b_f(\ma Q\hat{\ve g}))\times (\ma Q\hat{\ve g})
    \label{eq:gnew}
\end{equation}

are equivalent to the transformed equations of motion for the original state $\hat{\ve g}$
\begin{equation}
    \ma Q\frac{d}{dt}(\hat{\ve g})=\ma Q(-(\ma b_f \hat{\ve g})\times \hat{\ve g})
    \label{eq:gtemp}
\end{equation}
Here $\ma Q$ is an orthogonal matrix representing some combination of  rotation and reflection in 3 dimensions. To check the symmetry under $\ma Q$, we can use the identity $\ma Q (\ve v \times \ve u) = \det(\ma Q) (\ma Q\ve v) \times (\ma Q\ve u)$ for vectors $\ve u$ and $\ve v$ to bring $\ma Q$ into the cross product in (\ref{eq:gtemp}) and produce 
\begin{equation}
    \frac{d}{dt}(\ma Q\hat{\ve g})=-\det(\ma Q)(\ma Q\ma b_f \ma Q^{-1}(\ma Q\hat{\ve g}))\times (\ma Q\hat{\ve g})
    \label{eq:dQgdt}
\end{equation}
The dynamics are symmetric under $\ma Q$ if (\ref{eq:gnew}) and (\ref{eq:dQgdt}) are equivalent, which is satisfied when $\ma b_f =\det(\ma Q)\ma Q \ma b_f \ma Q^{-1}$ ~\citep{brenner1964stokes,fries2017angular,collins_lord_2021,sundberg2026}. In some cases, the dynamics can be symmetric with time reversed if $\ma b_f =-\det(\ma Q)\ma Q \ma b_f \ma Q^{-1}$. To capture both possibilities, we include a factor of $\tau$ defined by
\begin{equation}
\tau=
\begin{cases}
-1 & \text{if } t \mapsto -t\\[4pt]
+1 & \text{otherwise}
\end{cases}
\end{equation}
and the full symmetry condition on $\ma Q$ can be expressed as
\begin{equation}
     \tau \ma b_f =\det(\ma Q)\ma Q \ma b_f \ma Q^{-1} 
    \label{transf}
\end{equation}

For $\ma Q$ to be a dynamical symmetry of the particle, (\ref{transf}) must be satisfied either forwards ($\tau=1$) or backwards ($\tau=-1$) in time. Possible combinations of $\ma Q$ and $\tau$ which satisfy (\ref{transf}) can be classified by the corresponding symmetry of the equations of motion~(\ref{eq:eom}):

\begin{enumerate}
    \item \{$\det(\ma Q) = +1$,$\tau =+1$\} Pure rotation.

    \item \{$\det(\ma Q) = -1$,$\tau =+1$\} Reflection, and an optional rotation.

    \item \{$\det(\ma Q) = +1$,$\tau =-1$\} Pure rotation + time-reversal.
    
    \item \{$\det(\ma Q) = -1$,$\tau =-1$\} Reflection + time-reversal (a parity time (PT) transformation), and an optional rotation.
\end{enumerate}

We now ask: for a given particle and its corresponding $\ma{b}_f$, how can we identify nontrivial operations $\ma{Q}$ that satisfy (\ref{transf})?  
A natural place to start is with $\ma{Q}$ that are point group symmetries of the particle shape. Each element of the group leaves $\ma{b}_f$ invariant and is therefore also a dynamical symmetry (cases (i) and (ii)).   

Transformations $\ma{Q}$ that send $\ma{b}_f \mapsto -\ma{b}_f$ become dynamical symmetries when combined with time reversal (cases (iii) and (iv)). 
As an example, consider a boat propeller with three or more symmetric blades. This particle changes handedness under a reflection across a plane normal to its rotation axis. The reflection switches the signs of all eigenvalues of $\ma{b}_f$. The symmetry condition is not satisfied under forward time evolution, but becomes valid with time reversal. The angular dynamics are then PT symmetric (case (iv)), meaning that the dynamics of a propeller are indistinguishable from those of its appropriate mirror image evolving backward in time.

For some problems, it can be helpful to split $\ma b_f$ according to the translation theorem (\ref{b_trans}) and rewrite the symmetry condition on $\ma Q$ and $\tau$ (\ref{transf}) as:
\begin{equation}
    \tau\ma b_{m} + \tau(\ma c\times\ve r) = \det(\ma Q)\ma Q\ma b_{m}\ma Q^{-1} + \det(\ma Q)\ma Q (\ma c\times\ve r) \ma Q^{-1}
    \label{eq:SymExttemp}
\end{equation}\\
To separate the role of $\ma c$ and $\ve r$ in constraining $\ma Q$, we can rewrite (\ref{eq:SymExttemp}) as (see Appendix \ref{ortho_translationtheorem_appendix} for details):
\begin{equation}
    \tau\ma b_{m} + \tau(\ma c\times\ve r) = \det(\ma Q)\ma Q\ma b_{m}\ma Q^{-1} + (\ma Q \ma c \ma Q^{-1})\times(\ma Q\ve r)
    \label{eq:SymExt}
\end{equation}
In this form, the symmetry group of a general particle can be understood as the symmetry group of $\ma b_m$, minus those symmetries which are broken by $\ma c$ and $\ve r$. It is therefore important to understand the constraints associated with $\ma b_m$ alone, which are exactly the dynamical symmetries of a cocentered particle with $\ve r = 0$.

\subsection{Cocentered Particles}

For cocentered particles ($\ve r=0$), the orientation dynamics are governed solely by $\ma b_m$ which is determined by particle shape. In addition to particle shape symmetries, all cocentered particles inherit constraints arising from $\ma b_m$ being a symmetric pseudo-tensor which prevent them from converging to a steady orientation, instead tracing closed orbits in orientation space. All cocentered particles fall into one of four classes based on the eigenvalues of $\ma b_m$: where $\ma b_m$ is null, isotropic, axisymmetric, or triaxial.

\subsubsection{Cocentered with \ma b = $0$}
Cocentered particles with $\ma b_m = 0$, such as spheres and ellipsoids cannot reorient with respect to their driving in Stokes flow.
They sediment at some angle relative to gravity determined by the anisotropy of the translational drag tensor $\ma a$. 

\subsubsection{Cocentered with isotropic \ma b}
When the translation-rotation coupling tensor is isotropic, from (\ref{eq:g}) we have $\frac{d}{dt} \hat{\ve g} \propto \hat{\ve g} \times \hat{\ve g} = 0$, and so every orientation is a fixed point. Such particles simply rotate about the gravity vector.  Known particles with isotropic $\ma b$ are isotropic helicoids~\citep{Kelvin} and have very weak translation-rotation coupling~\citep{collins_lord_2021}. They fall straight down with a possible slow rotation.   More complex spatial trajectories may occur if $\ma a$ is anisotropic, but no known particles have isotropic $\ma b$ with anisotropic $\ma a$.

\subsubsection{Cocentered with axisymmetric \ma b}
Cocentered particles with axisymmetric $\ma b_m$ include helical ribbons with a special length~\citep{huseby2025helical} or symmetric propellers with more than two blades.  The continuous rotational symmetry renders the dynamics one-dimensional, so that the gravity vector $\hat{\ve g}(t)$ traces circular orbits around the particle's symmetry axis. 
In lab coordinates, this corresponds to a particle that can spin around its axis of rotational symmetry and around $\hat{\ve g}$, with a fixed angle between the two axes.
Such particles can sediment vertically or trace a helix around the gravity vector.  

\subsubsection{Cocentered with triaxial \ma b}
Triaxial $\ma b_m$ is the minimal symmetry case for a cocentered particle. Typical helical ribbons are triaxial.  
Since $\ma b_m$ is always symmetric, it has 3 real eigenvectors which correspond to 6 fixed points of (\ref{eq:g}) where the particle spins steadily around gravity. There are 3 planes of PT symmetry, one orthogonal to each $\ma b_m$ eigenvector. Each of the 6 orientation fixed points lies at the intersection of two such PT symmetry planes, one of which is sufficient to restrict the dynamics to be reversible across that plane and the fixed points to be either saddles or centers. So, in general, cocentered particles cannot converge to a steady orientation.

The total topological index~\citep{Strogatz} of singularities on the sphere must be equal to two, 
which, combined with the 6 fixed points, requires 4 centers and 2 saddles. The saddles lie along the eigenvector associated with the intermediate eigenvalue of $\ma b_m$ (see Appendix~\ref{fpstability1}). So, by symmetry considerations alone, we know that cocentered particles with triaxial translation-rotation coupling only trace closed orbits around 4 centers separated by 2 saddles.  Such particles trace spirograph trajectories as they sediment~\citep{huseby2025helical}.  The orientation dynamics of an example cocentered triaxial particle is shown in Figure~\ref{fig:cocentered}, in terms of the gravity vector's trajectory in body coordinates and the plane formed by the two relevant Euler angles.

\subsubsection{Special Case: Cocentered with two-dimensional $\ma b$}
When $\ma b_m$ has one eigenvalue that is zero, the translation-rotation coupling is two-dimensional. Known cases are non-chiral particles for which the reflection symmetry forces one eigenvalue to be zero and the other two  to be an equal and opposite pair such as the two bladed impeller or perpendicular two-bladed propeller~\citep{Happel}, and the oloid~\citep{flapper2025}.  Another two-dimensional cocentered particle can be made from a bent disk~\citep{miara_dynamics_2024} or a curved fibre~\citep{candalier2025curvedfibres} with density distribution modified to move the center of force to the center of mobility.  The sedimentation dynamics for these cases matches the tri-axial case above except that the saddle point at the zero eigenvector is not associated with a steady rotation around gravity.

Non-cocentered particles with two-dimensional $\ma b_m$ have been studied extensively including the asymmetric two bladed propeller~\citep{kim2013microhydrodynamics}, bent disks~\citep{miara_dynamics_2024} and more general shapes with two reflection symmetries~\citep{joshi2025sedimentation}. These can be understood as a case with center of force offset from the center of  a cocentered reference particle which is tri-axial and two-dimensional. When the offset is along the intermediate axis of $\ma b_m$ and small, we will show that  the dynamics can remain reversible with closed orbits.   

\subsection{Connecting cocentered particles with the general case}

Particles with fewer geometric symmetries, the typical case, are unlikely to be cocentered and tend to have non-zero \textbf{r}.   We have already considered the $\ve r = 0$ case and seen that such particles do not have attracting orientations. Conversely, when $|\ve r|\xrightarrow{} \infty$ the gravitational torque diverges and sedimenting particles reorient such that $\ve r$ and $\hat{\ve g}$ are aligned, at a unique orientation. The transition between these two limits, at intermediate $|\ve r|$, is not well understood and is our main focus.

Behavior in the intermediate $|\ve r|$ regime can be understood by identifying a cocentered reference particle and asking which of its symmetries are broken by the introduction of $\ve r$ and $\ma c$. More generally, when $\ve r$ preserves no symmetries, we can represent each particle as a point in the three dimensional parameter space defined by $\ve r$ and gain qualitative insights about the behaviors available to non-cocentered particles.

It is useful to distinguish cases where the eigenframe of $\ma c$ is the same as the eigenframe of $\ma b_m$ from cases where they are different. For helical ribbons, the eigenframes of $\ma b_m$ and $\ma c$ are always the same.  This is the simplest case to understand and will be the focus of this paper. In the axisymmetric and triaxial cases listed above, rotation between the eigenframes of $\ma c$ and $\ma b_m$ is possible, for example, the bent disk and two bladed propellers have their eigenframes rotated by $\pi/4$. We address differences associated with this possible rotation when relevant.

\section{Methods}

\subsection{Helical Ribbons}

\cite{huseby2025helical} introduced helical ribbons as simple cocentered particles that can be precisely fabricated and tracked in three-dimensions during sedimentation. Helical ribbons have three orthogonal $\pi$ rotational symmetries, which ensure that the centers of resistance, mobility, and buoyancy are all coincident, since any rotational symmetry fixes all centers associated with the particle's shape to lie along the rotation axis. Helical ribbons with homogeneous density, explored previously, have their center of mass at the center of buoyancy and are therefore cocentered.

Here we extend previous cocentered experiments by measuring the dynamics of helical ribbons with centers of mass offset from their geometric center. Experimentally, we limit center of mass offsets to lie along the three principal axes of the particle.

\begin{figure}
    \centering
    \includegraphics[width=0.8\linewidth]{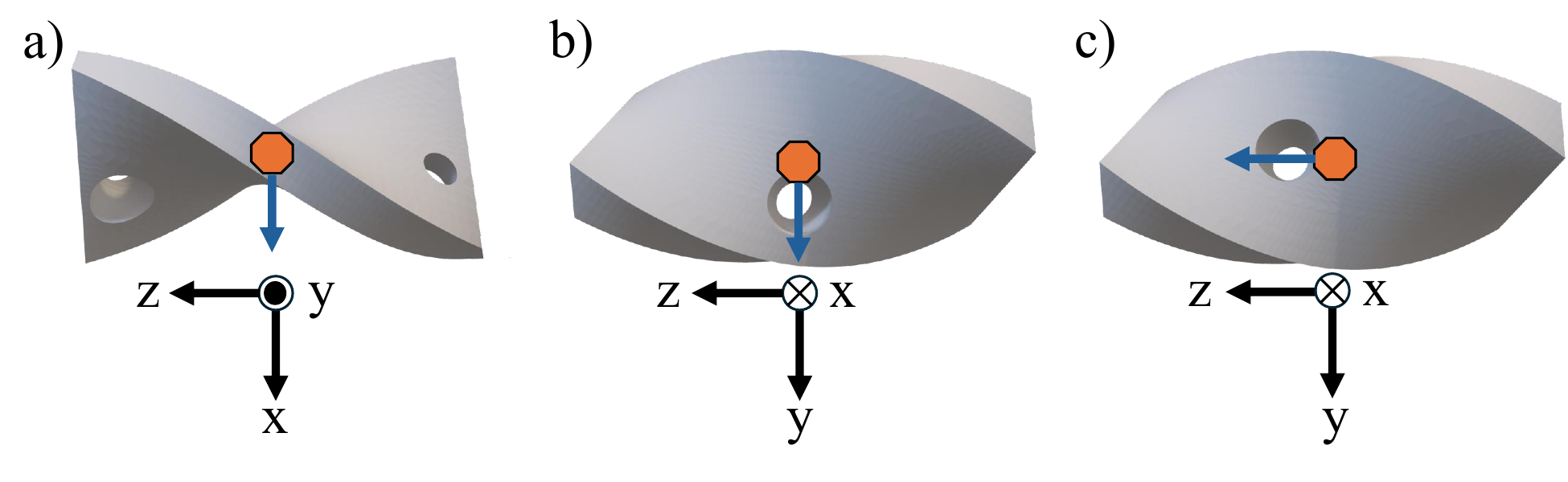}
    \caption{Helical ribbons with holes where metal spheres are epoxied to move the center of mass (COM) along principal axes.  (a) COM offset in $x$ direction which is the intermediate axis (labeled by eigenvalues of $\ma b_m$).  (b) COM offset in $y$ direction which is the minor axis.   (c) COM offset in $z$ direction which is the major axis.}
    \label{fig:com_offset}
\end{figure}

\subsection{Particle Manufacturing}
Particles are fabricated with a Formlabs Form 2 3D printer with Rigid 4000 resin.  To move the center of mass we create CAD designs that have holes at different locations in the particles as shown in Figure~\ref{fig:com_offset}. Spheres made of aluminum, steel, or tungsten with diameters of $1$mm are epoxied into the holes to create particles with nearly identical geometric shape, and thereby mobility tensors ($\ma a$, $\ma b_m$, $\ma c$), but different centers of mass. A single metal sphere can move the center of mass along the $y$ or $z$ axes, while the $x$ axis is the thickness direction at the center of the ribbon and so two spheres are placed toward the corners of the helical ribbon to move the center of mass in the $x$ direction.  Since the geometry is essentially unchanged, the center of buoyancy does not move significantly.

Since sedimentation is very sensitive to center of mass offsets, we find that we have to correct for an intrinsic density inhomogeneity produced by the Formlabs printer. In \cite{huseby2025helical}, special particles fabricated by Boston Microfabrication were used to avoid this density inhomogeneity.  Here we measure particles with the bead position shifted along each principal axis and use the sedimentation data to measure the deviation from cocentered dynamics.  A linear fit provides the location where the bead would exactly correct for the printer inhomogeneity, and then the actual center of mass of each particle can be computed geometrically.  Details are included in Appendix~\ref{Exp_appendix}. 

Using the results from ~\citet{huseby2025helical} we choose the $3\pi/4$ helical ribbon since it has a tri-axial $\ma b_m$ with large values for all three eigenvalues. Particles here have length $L=7.5$mm, $2 \pi L/s= 3\pi/4$, where $s=20$mm is the length over which the helix twists through a full rotation, width $W=4$~mm, and thickness $T=1$~mm.  The thickness differs from ~\citet{huseby2025helical} to accommodate the metal spheres to move the center of mass.

Particles are dropped in silicon oil with viscosity 1000 cSt and are tracked using video images from three nearly orthogonal cameras with pulsed LED backlighting using methods described in~\citet{oehmke_spinning_2021} and \citet{huseby2025helical}.     

\subsection{Numerical simulation of mobility tensors} \label{sim_mobility_tensors}

We numerically calculate the mobility tensor elements for a helical ribbon using the Immersed Boundary Method (IBM). In this method, the incompressible Navier-Stokes equations are solved on a fixed Cartesian grid while the particle surface is resolved using a set of Lagrangian markers. The liquid and solid phases are coupled through a forcing term added to the momentum equation, which enforces the no-slip boundary condition on the particle surface. 

The advection and diffusion terms in the momentum equation are discretized using a two-step Adams-Bashforth scheme and the Crank-Nicolson method, respectively. The pressure field is obtained by solving a Poisson equation to enforce incompressibility. The translation and rotation of the particle are determined through the Newton-Euler equations. The dynamics of the particle and liquid phase are advanced in time using a three-step Runge-Kutta method. This IBM framework has been extensively validated and widely applied in various studies \citep{verzicco1996finite, breugem2012second, detullio2016moving, witkamp2024low, wang2025inertial}.  

The mobility tensors are estimated using the inverse formulation of~\eqref{eqft},
\begin{equation}
\begin{bmatrix} \ve f \\ \ve \tau\end{bmatrix}  
= \mu 
\begin{bmatrix}\ma a' & {\ma b'}^{\top}  \\\ma b' & \ma c'  \end{bmatrix}^{-1}
\begin{bmatrix}\ve v \\\ve \omega\end{bmatrix},
\label{resmat}
\end{equation}
where the 6$\times$6 matrix on the right hand side of the equation represents the grand resistance tensor. 

The grand resistance tensor is determined numerically using six separate simulations. In each simulation, either a translational or rotational velocity component of the particle is prescribed along one coordinate axis in quiescent fluid, while all the other particle velocity components are set to zero. The resulting hydrodynamic forces and torques exerted on the particle then define one column of the grand resistance tensor. The mobility tensors $\ma a$, $\ma b$, and $\ma c$ are subsequently retrieved by taking the inverse of the grand resistance tensor. This procedure for evaluating the mobility tensors has been validated and used for several particle geometries, and the relative error in the non-zero terms of the tensor was found to be approximately 2\% for a sphere \citep{witkamp2024low}. 

The numerically obtained mobility tensors for the helical ribbon are 
\begin{align}
6\pi r_h \; \ma a &=
\begin{bmatrix}
0.91 & 0 & 0\\
0 & 0.95 & 0 \\
0 & 0 & 1.13
\end{bmatrix} 
\label{eq:simvalues_a}\\[1ex]
r_h^2 \; \ma b_m &=
\begin{bmatrix}
1.67 \times 10^{-3} & 0 & 0\\
0 & -3.44 \times 10^{-3}  & 0 \\
0 & 0 & 5.01 \times 10^{-3} 
\end{bmatrix} 
\label{eq:simvalues_b}\\[1ex]
8 \pi r^3_h \; \ma c &=
\begin{bmatrix}
1.13 & 0 & 0\\
0 & 1.12 & 0 \\
0 & 0 & 2.56
\end{bmatrix}
\label{eq:simvalues_c}
\end{align}
where we normalize by the hydrodynamic radius, $r_h = 1/(2 \pi \; \mathrm{trace} (\ma a))=3.11$ mm, in such a way that the normalized \ma a and \ma c tensors are identity matrices for a spherical particle. 

\subsection{Numerical simulation of sedimentation dynamics} \label{sim_sedimentation_dynamics}

We numerically simulate (\ref{eq:eom}) to obtain the dynamics of sedimenting helical ribbons with the mobility tensor values in Eqs. (\ref{eq:simvalues_a}-\ref{eq:simvalues_c}) using a standard ordinary differential equation solver (Matlab ODE45). Equation (\ref{b_trans}) is used to convert the translation-rotation coupling tensor to the center of force. Animations of the phase space dynamics are visualized in movies available in Appendix~\ref{supplementary_videos}.

\section{Results: Uniaxial Offsets}
We start by offsetting the center of mass of the helical ribbon along each of its principal axes. These axes are eigenvectors of  $\ma b_m$ and are fixed by the three axes with symmetry under rotation by $\pi$. We compare experimental phase space trajectories to predicted bifurcation diagrams, computed by calculating the Jacobian around each of the real eigenvectors of $\ma b_f$ where the fixed points reside. 

\subsection{Experiments}

\begin{figure*}
    \centering
    \includegraphics[width=1\linewidth]{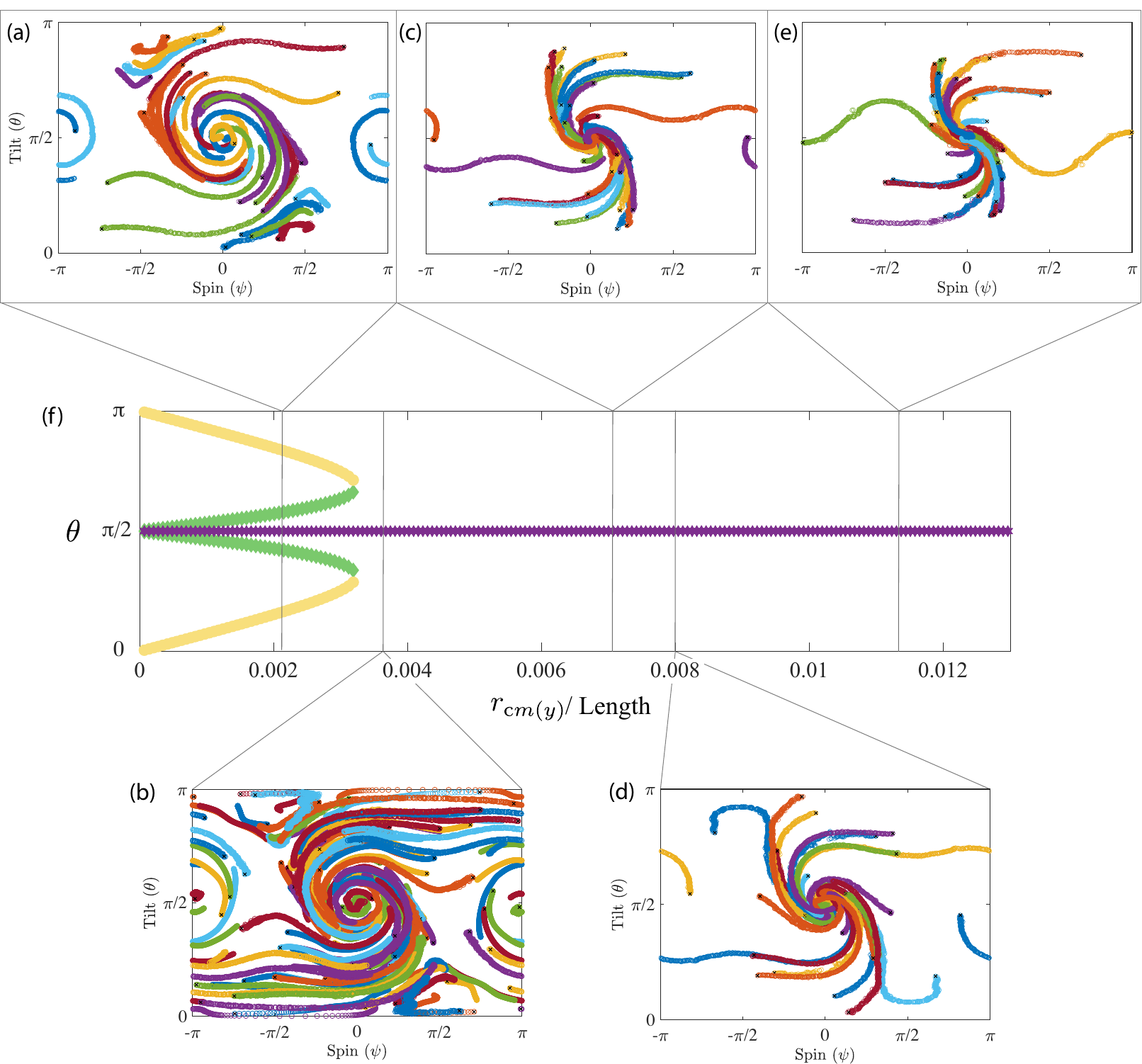}
     \caption{Increasing $y$ (minor axis) offset of the center of mass.  (a) through (e) show experimental phase diagrams as the center of mass moves along the $y$ axis.  (f) shows a bifurcation diagram of the tilt ($\theta$) position of the fixed points from numerical simulations.  Saddles are green and centers are yellow.  Vertical lines in (f) show the center of mass offset of each experimental data set.  }
    \label{fig:bifminor}
\end{figure*}

Figure~\ref{fig:bifminor}(a-e) shows experimental measurements of a helical ribbon's orientation dynamics with the center of mass offset along the particle's minor axis ($\hat{y}$). The offset $\ve r$ is co-linear with the centers at $\theta=\pi/2$ and $\psi=0,\pm\pi$, so their positions are fixed as $\ve r$ increases and they become stable and unstable spirals respectively.  

The other two centers are initially along the major axis ($\hat{z}$), starting at $\theta = 0$ and $\theta = \pi/2$. As the center of force moves, these fixed points migrate towards and eventually annihilate with the two saddles. After the final bifurcation, only the two spirals along $\ve r$ remain, with the stable one attracting all orientation trajectories. We refer to this as a 6 to 2 bifurcation in which 6 fixed points become 2.  The phase space dynamics can be clearly visualized in movies linked from Appendix~\ref{supplementary_videos} which demonstrate that the center fixed points at $\psi = \pm \pi/2$ that migrate and annihilate remain centers until the bifurcation.

The bifurcation diagram shown in  Figure~\ref{fig:bifminor}(f) clearly shows the saddle-node bifurcation that occurs when the center of force moves along the minor axis. Note, the magnitude of $\ve r$ where the final bifurcation occurs for the numerically simulated trajectories is 0.0032$L$ which is 26 $\mu$m, a tiny fraction of the size of the ribbon. The experimental data suggests the bifurcation occurs at somewhat larger offset, somewhere between $0.0036L$ and $0.0070L$.  This discrepancy is likely a result of either differences in the mobility tensors between experiments and the immersed boundary simulations or uncertainties in the measurement of the center of mass of the particles.

\begin{figure*}
    \centering
   \includegraphics[width=1\linewidth]{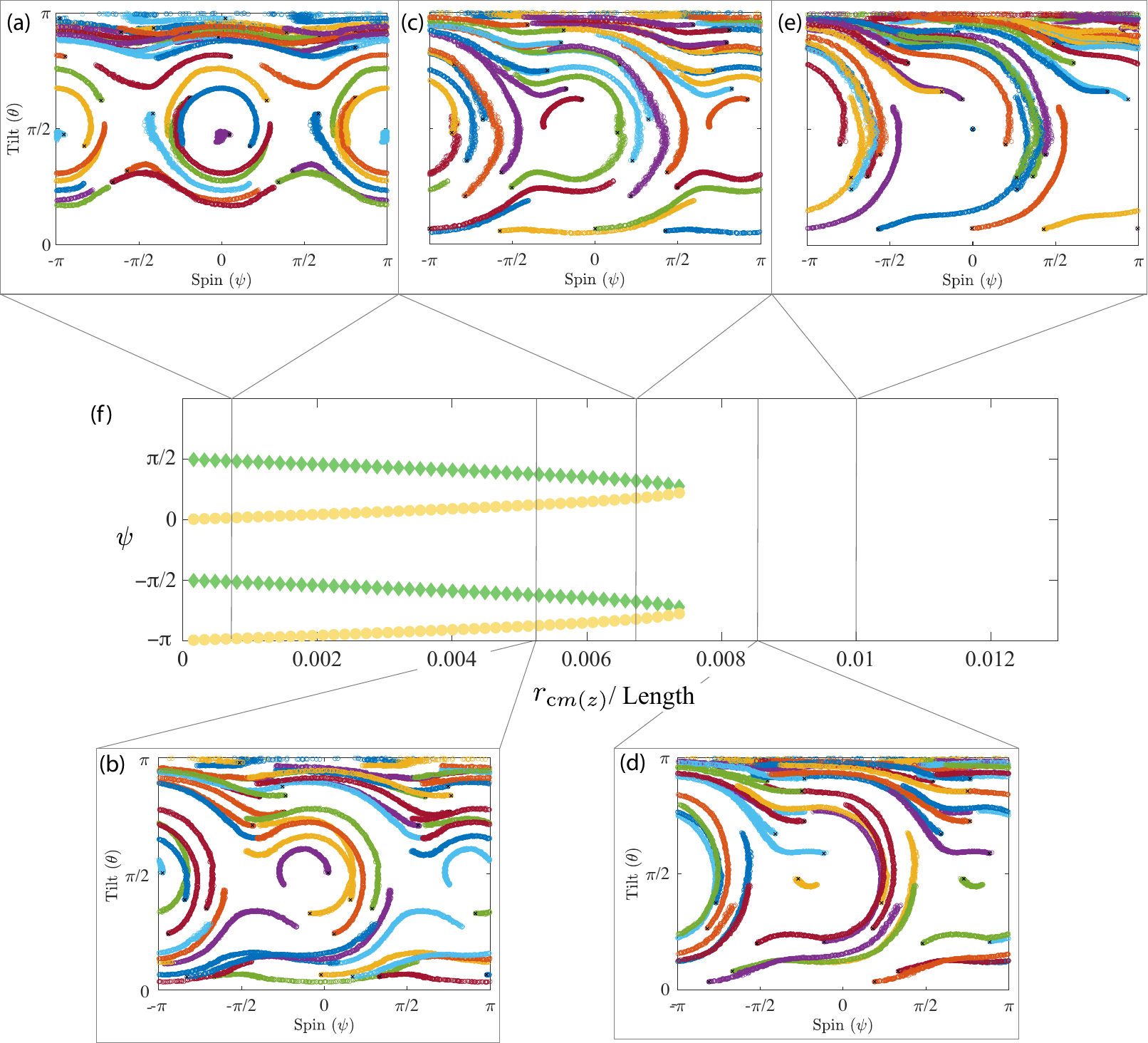}
    \caption{Increasing $z$ offset of the center of mass.  (a) through (e) show experimental phase diagrams as the center of mass moves along the $z$ axis.  (f) shows a bifurcation diagram of the spin ($\psi$) position of the fixed points from numerical simulations.  Saddles are green and centers are yellow.  Vertical lines in (f) show the center of mass offset of each experimental data set.}
    \label{fig:bifmajor}
\end{figure*}

Offsetting the center of force along the major axis ($\hat{z}$) of the helical ribbon produces the experimental phase spaces and bifurcation diagram pictured in Figure~\ref{fig:bifmajor}. The qualitative behavior is the same as that seen for offsets along the minor axis. Now, the centers at $\theta = 0$ and $\theta = \pi$, where the ribbon falls with its long axis vertical, become stable and unstable spirals. The centers at $\theta=\pi/2$ remain centers, migrate in $\psi$, and annihilate with the saddles. It is harder to see the spirals in the Euler angle plots since the fixed points are at the poles. In the experimental data, we can see more clearly in 
Figure~\ref{fig:bifmajor} than in Figure~\ref{fig:bifminor} that some closed orbits remain up until the 6 to 2 bifurcation. Here the bifurcation occurs for center of mass offsets more than a factor of two larger than for the minor ($\hat{y}$) axis, but still much smaller than any dimension of the ribbon itself. 

\begin{figure*}
    \centering
     \includegraphics[width=1\linewidth]{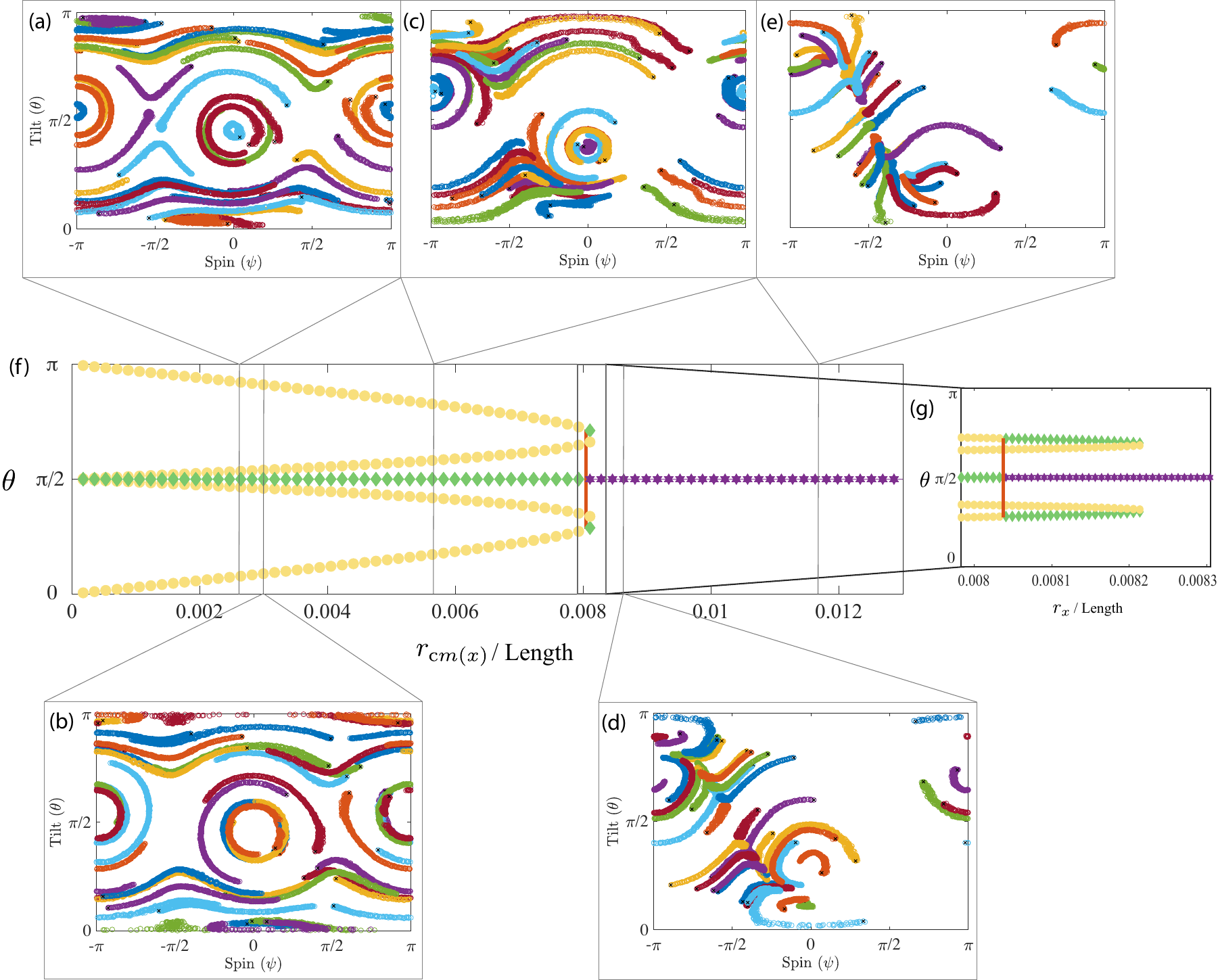}
    \caption{Increasing offset of the center of mass along the intermediate ($x$) axis.  (a) through (e) show experimental phase diagrams as the center of mass moves along the $x$ axis.  (f) shows a bifurcation diagram of the tilt ($\theta$) position of the fixed points from numerical simulations.  Saddles are green, centers are yellow, nodes are in purple and the line of nodes is orange. (g) shows a magnified view of the line of nodes bifurcation and the saddle node bifurcation.  Vertical lines in (f) show the center of mass offset of each experimental data set.}
    \label{fig:bifintermediate}
\end{figure*}

Offsetting the center of mass of a helical ribbon along its intermediate axis ($\hat{x}$) results in a more complicated sequence of bifurcations than the cases above. Experimental phase spaces and bifurcation diagram for this case are shown in Figure~\ref{fig:bifintermediate}.  For small offsets, the centers migrate towards one another in pairs while the saddle points remain fixed, because they are co-linear with $\ve r$. In the experimental phase space plots in Figure~\ref{fig:bifintermediate}, the centers then appear to annihilate, leaving behind one stable and one unstable node. However, such a bifurcation is topologically forbidden: since centers each carry an index of +1, they cannot annihilate with one another while conserving the total topological charge of the space. Upon closer numerical inspection, there is a line of nodes bifurcation at which the saddles transform into nodes and two of the centers transform into saddles, which conserves topological charge. Then, the new saddles and the remaining centers annihilate in a saddle-node bifurcation which is the familiar 6 to 2 bifurcation.  The inset in Figure~\ref{fig:bifintermediate}(f) shows this sequence of bifurcations which occurs over a very small range of center of mass offsets.

The critical value $r_x^*$ at which the saddle bifurcates into a line of nodes can be found by calculating the Jacobian

\begin{equation}
    J = \nabla_g\frac{d\hat{\ve g}}{dt}
    \label{Jacobian}
\end{equation}

for $\hat{\ve g}$ along $\hat{x}$, and setting $\det \ma J = 0$, to identify the switch from a saddle to a node, which gives
\begin{eqnarray}
    r^*_x = \pm\sqrt{\frac{(b_3-b_2)(b_2-b_1)}{c_1c_3}}
    \label{eq:surfscale}
\end{eqnarray}
where $b_i$ and $c_i$ are the $i^{th}$ eigenvalues of $\ma b_m$ and $\ma c$ respectively, ordered so that $b_1\leq b_2\leq b_3$.  See Appendix~\ref{fpstability2} for details.

This expression shows that particles with less rotational drag, larger $\ma c$, are more sensitive to small center of mass offsets. Additionally, it highlights a useful length scale, $|\ma b_m|/|\ma c|$, which is a property of a shape and sets the scale of center of force offsets at which bifurcations typically occur. This length scale is typically much less than the hydrodynamic radius. Typical shapes have weak translation-rotation coupling relative to their rotational drag, making $|\ma b_m|/|\ma c|$ small, which explains why most particles are extremely sensitive to density inhomogeneity.  

See the Appendices~\ref{fpstability1} and~\ref{fpstability2} for a more complete discussion of the stability around each fixed point. The mathematical reason that the intermediate axis is special is very closely related to the reason for the intermediate axis theorem in rotations of a solid body in classical mechanics.
Videos illustrating each of the above bifurcations, as we vary the center of mass position are linked in Appendix~\ref{supplementary_videos}.

\subsection{Theory for Uniaxial Offsets}

The three PT symmetries of a cocentered particles forbid formation of spirals, migration of fixed points and bifurcations which are observed experimentally. So we need to understand which PT symmetries are broken by a center of mass offset.  We sketch an intuitive argument and then show how this follows from the application of (\ref{eq:SymExt}). 

When the center of mass moves, the associated torque must switch signs like $\ma b_m$ under reflection for the corresponding PT symmetry to be preserved. This is only possible when the offset is normal to one of the original reflection planes, otherwise the torque does not switch signs correctly and the dynamics are no longer time-reversal symmetric. So, the center of mass offset breaks all PT symmetry planes in which it has a component, leaving at most one PT symmetry behind. Mathematically, this follows from (\ref{eq:SymExt}). Start by assuming $\ma Q$ is a reflection associated with one of the particle's original cocentered PT symmetries, which are each normal to one of the eigenvectors of $\ma b_m$, and ask if it remains a PT symmetry of the dynamics when $\ve r \neq 0$. Since $\tau=-1$ and det$(\ma Q)=-1$, we can rewrite equation (\ref{eq:SymExt}) as

\begin{equation}
   \ma c\times \ve r = -(\ma Q \ma c \ma Q^{-1})\times(\ma Q\ve r)
   \label{eq:temp3}
\end{equation}

In the case of helical ribbons, where $\ma b$ and $\ma c$ share an eigenframe, $\ma Q$ is just a reflection normal to a $\ma c$ eigenvector and (\ref{eq:temp3}) reduces to

\begin{equation}
   \ma c\times \ve r = \ma c\times(-\ma Q\ve r)
   \label{eq:temp4}
\end{equation}
which is only satisfied when $\ve r$ is normal to $\ma Q$, so that $\ma Q \ve r = -\ve r$. So, the reversibility from a PT symmetry is destroyed whenever $\ve r$ has a component in a plane of reflection symmetry, consistent with experimental observations in Figures~\ref{fig:bifminor}-\ref{fig:bifintermediate}.

For PT symmetries associated with eigenvectors of \ma b that are not also eigenvectors of \ma c, reversibility is typically broken for any $\ve r$.    The reversible dynamics of the bent disk~\citep{miara_dynamics_2024} and the curved wires~\citep{candalier2025curvedfibres} can be understood in this framework.  Although \ma b and \ma c do not have the same eigenframe, they do share one eigenvector.  The two reflection symmetries of the particle ensure that this shared eigenvector is the same axis along which the center of mass is offset from the center of mobility, so reversibility is maintained. Because the shared eigenvector is the intermediate axis of $\ma b_m$, all orbits are closed.  Note that the preserved PT symmetry is not associated with one of the reflections of the particle geometry, and is in fact perpendicular to both. 

In summary, cocentered particles have 3 PT symmetries along eigenvectors of $\ma b_m$. The introduction of a nonzero offset $\ve r$ breaks all PT symmetries except for the special case where $\ve r$ is normal to one such reflection plane which is also an eigenvector of $\ma c$, in which case that PT symmetry is preserved.

\begin{figure*}
    \centering
    \includegraphics[width=1\linewidth]{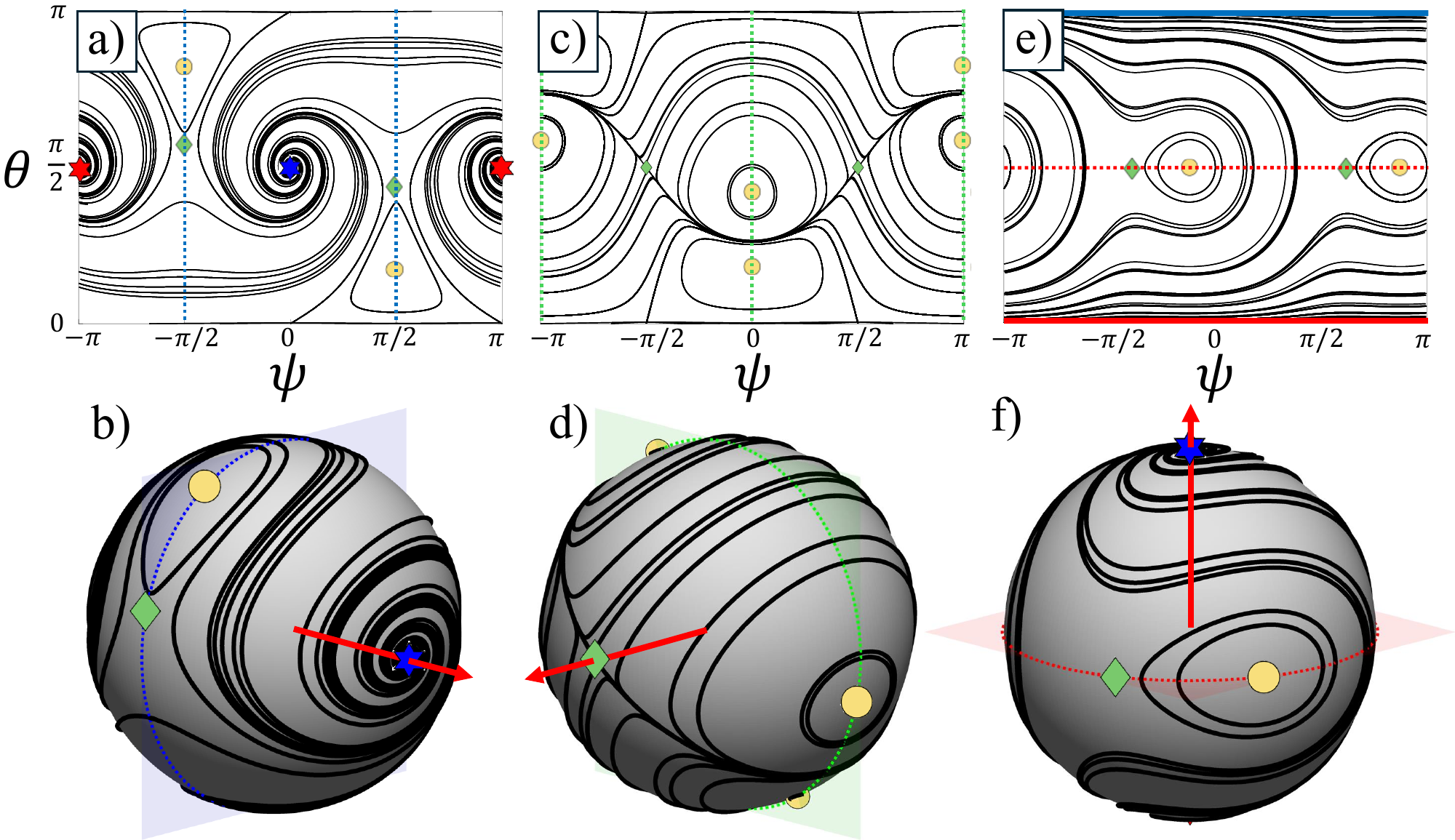}
    \caption{Computationally generated phase spaces for a helical ribbon a with small center of mass offset along its (a-b) minor ($\hat{y}$) (c-d) intermediate ($\hat{x}$) (e-f) and major ($\hat{z}$) axes. The red arrows indicate offset direction $\ve r$. The translucent panes and dotted lines represent reflection planes with preserved PT symmetry. In (a) the blue stars are attracting spirals and the red stars are repelling spirals. In (e) the blue and red bars at $\theta=0$ and $\theta=\pi$ correspond to attracting and repelling spirals at the poles.  Animations of the dependence of (a), (c), and (e) on the magnitude of the center of mass offset are available at links in Appendix~\ref{supplementary_videos}.}
    \label{fig:symmetry}
\end{figure*}

Figure~\ref{fig:cocentered} illustrated the orientation dynamics of a cocentered particle, with closed orbits around 4 centers separated by 2 saddles. Figure \ref{fig:symmetry} shows a similar visualization, now with center of mass offset. In each case, the PT symmetry normal to $\ve r$ survives and the other two are broken.  When $\ve r$ lies along the major or minor axis of $\ma b_m$, as pictured in Figure~\ref{fig:symmetry}(a-b, e-f), the unbroken PT symmetry plane intersects the pair of saddles and centers orthogonal to $\ve r$. Their stability is fixed by reversibility over the remaining reflection plane, but all 4 are free to migrate along the great circle where the remaining reflection plane intersects the unit sphere. The remaining two fixed points, which lie along $\ve r$, are pinned to their position by the unbroken PT symmetry.  They can change their stability to become spirals since reflection plus time reversal turns an attracting spiral at one pole into a repelling spiral at the other pole. Closed orbits are preserved in a region around the last two centers until $\ve r$ is large enough that the centers and saddles annihilate.

When $\ve r$ lies along the intermediate axis of $\ma b_m$ shown  in Figure~\ref{fig:symmetry}(c-d), the remaining PT symmetry plane intersects all 4 centers.  The other two fixed points, which are pinned, are saddles.  The centers cannot become spirals without violating the reversibility constraint, while the saddles are topologically forbidden from swapping to spirals by themselves. So, there are non-zero values of $\ve r$ where all orbits remain closed. The centers are still free to migrate within the remaining reflection plane.

When the bifurcation from 6 to 2 fixed points occurs, the eigenvalues of $\ma b_f$ associated with the 4 annihilated fixed points become complex. In all cases the 2 saddle node bifurcations happen simultaneously, because the eigenvalues of $\ma b_f$ form a complex conjugate pair. So, regardless of the direction of $\ve r$, a globally stable spiral can only emerge through a 6 to 2 bifurcation. 

\section{Results: Arbitrary Offsets}

\subsection{Bifurcations for Arbitrary Center of Mass Offsets}

For center of mass offset along each principal axes of $\ma b_m$, we observe a 6 to 2 bifurcation at some critical displacement between the centers of mass and mobility.  Beyond the bifurcation, the particle converges to a single stable orientation. In experiments, we only moved the center of force along principal axes, but with numerical simulations we can explore arbitrary offsets. 

We can define a three dimensional space representing all possible displacement vectors $\ve r$ in the eigenframe of the $\ma b_m$ tensor.  Each value of $\ve r$ corresponds to a distinct particle with the same shape, but different density distribution.  In this space, there are surfaces where bifurcations occur. The dynamics of a sedimenting particle depend smoothly on $\ve r$, so the bifurcation surfaces will separate parameter space into discrete volumes with qualitatively distinct behavior. 

\begin{figure*}
    \centering
    \includegraphics[width=0.6\linewidth]{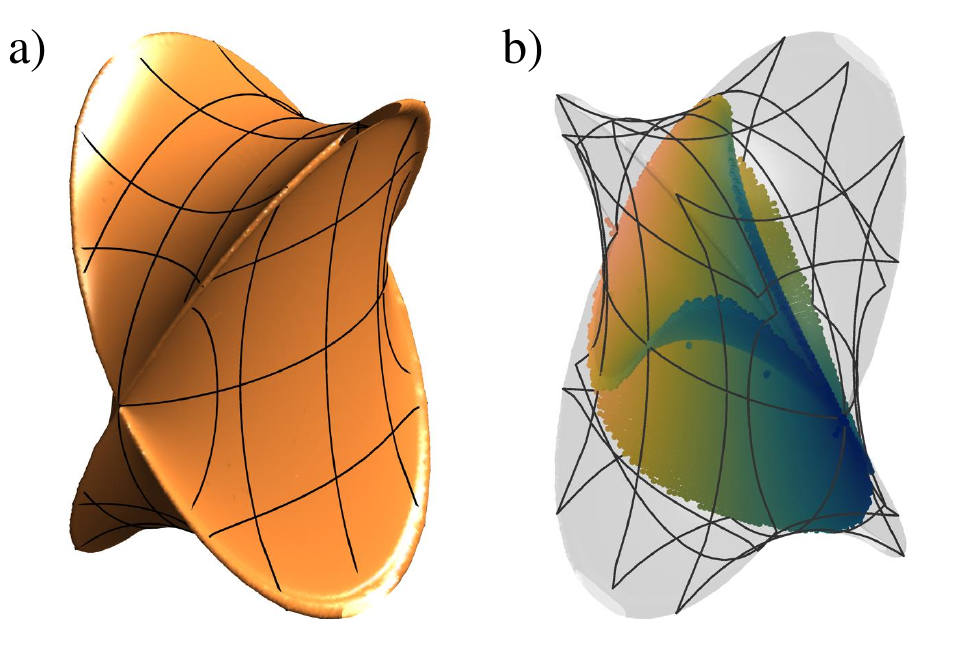}
    \caption{Surfaces defining bifurcations as a function of the center of mass position. (a) the alignment bifurcation surface where the dynamics switches from 6 to 2 fixed points.  Particles outside the surface have a single stable orientation. (b) includes the internal surface defined by the location of a Hopf bifurcation.}
    \label{fig:surface}
\end{figure*}

The bifurcation surface with the most significant change in sedimentation dynamics occurs where the double saddle-node bifurcation changes 6 fixed points into 2.  We call this surface the alignment bifurcation surface.  
We can determine the location of this surface by computing when two of $\ma b_f$'s eigenvalues become complex conjugates. Figure~\ref{fig:surface}(a) illustrates the alignment bifurcation surface using numerically simulated mobility tensors. Any particle whose center of force displacement lies outside this surface will converge to a unique orientation.

The alignment bifurcation surface displayed here, associated with a helical ribbon, is especially symmetric because $\ma b_m$ and $\ma c$ share an eigenframe. Two of the helical ribbon's $\pi$ rotation symmetries are sufficient to leave only 2 unique octants for all bifurcation surfaces, from which the rest can be reconstructed. Bifurcation surfaces of more complex particles, without the constraints upon $\ma c$'s eigenframe, tend to be stretched or rotated versions of the one found here. In all alignment bifurcation surfaces we have studied, there are lower dimensional structures on the bifurcation surface, typically four co-dimension 2 cusps which appear to meet at two co-dimension 3 points. 

The alignment bifurcation surface delineates a sharp transition from complicated, small $\ve r$, dynamics to the simple convergence exhibited at large $\ve r$. The bifurcation surface's shape and size encodes the full 3D sensitivity of a particle to density inhomogeneity. This transition between small and large $\ve r$ occurs at the bifurcation surface and the overall size of the surface scales with $|\ma b_m|/|\ma c|$.

\subsection{Complex Dynamics for Particles With Small Offsets}

Particles with center of mass offsets inside the bifurcation surface typically have complex dynamics with two attracting spirals, two repelling spirals, and two saddles. We find there are several other bifurcations that occur within the alignment bifurcation surface. To explore this behavior, we take a slice through the bifurcation surface, pictured in Figure~\ref{fig:map}, and vary $\ve r$ along trajectories within that slice of the center of force offset space.

\subsubsection{Limit Cycles}

\begin{figure*}
    \centering
    \includegraphics[width=0.75\linewidth]{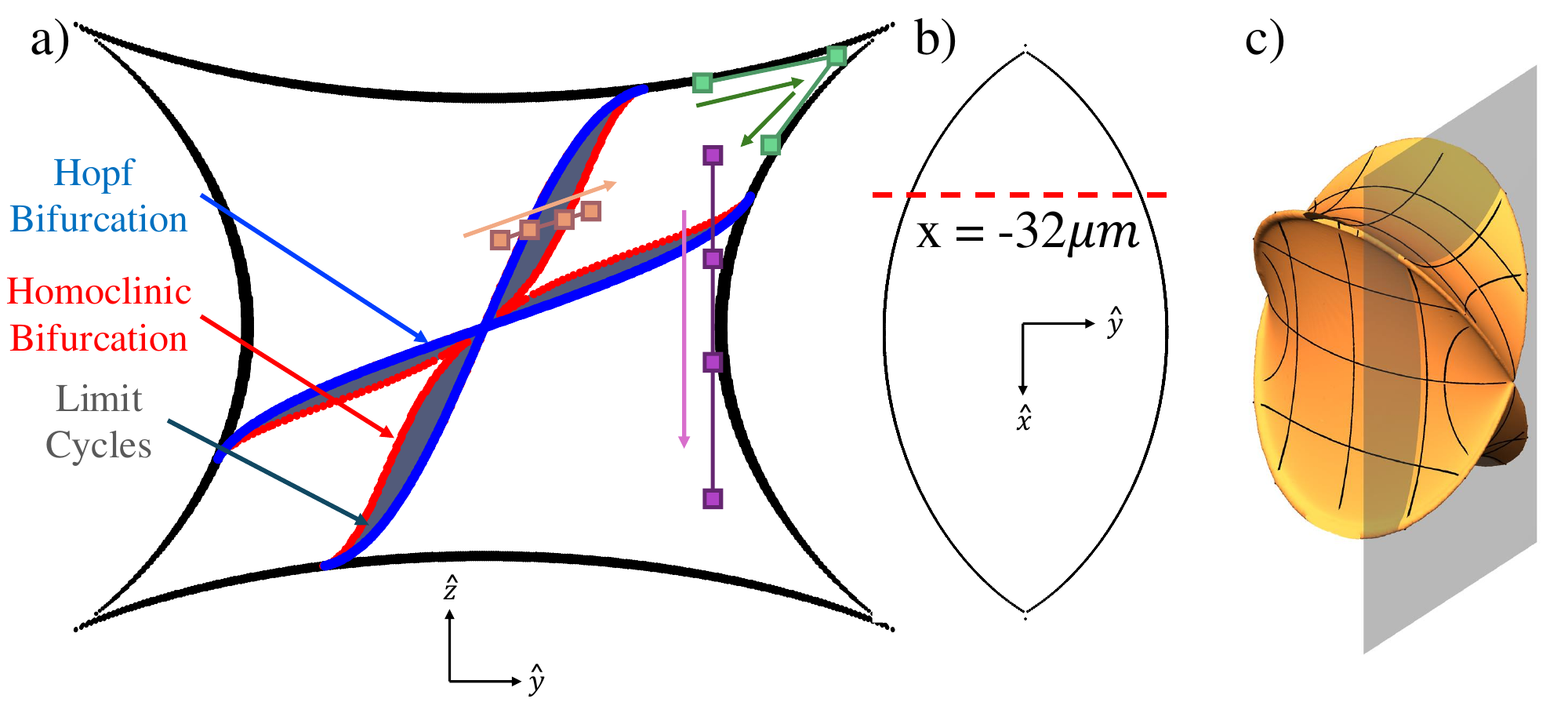}
    \caption{A schematic of dynamical regimes in a slice of the helical ribbon's bifurcation surface. (a) shows the location of homoclinic bifurcations (red), Hopf bifurcations (blue), and Limit Cycles (grey), as well as the series of specific examples explored in later figures (tan, green, and purple squares). (b) and (c) show the location of the cross section in the alignment bifurcation surface.}
    \label{fig:map}
\end{figure*}

\begin{figure*}
    \centering
    \includegraphics[width=1\linewidth]{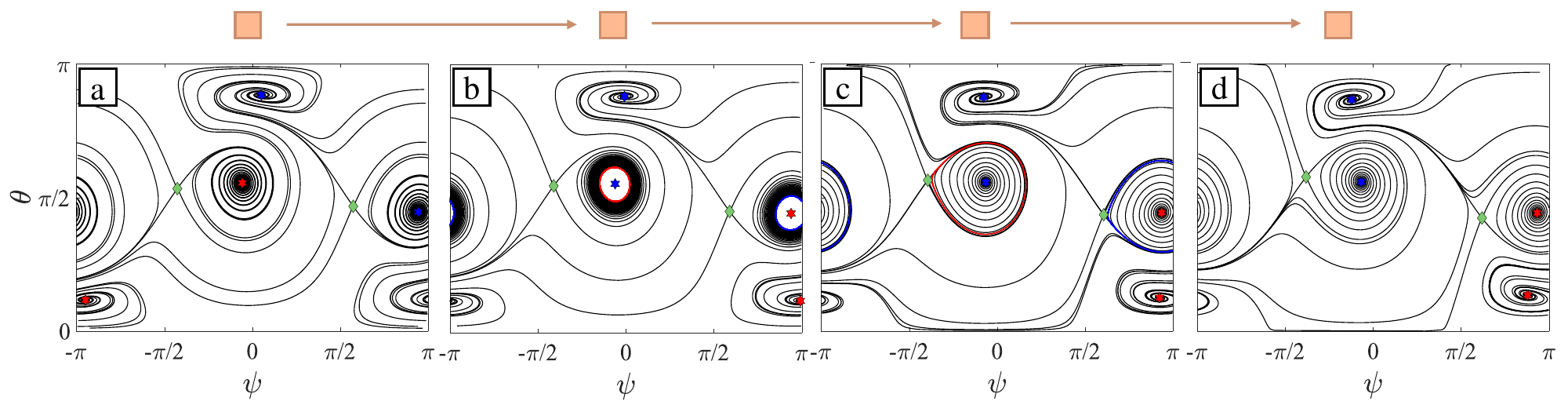}
    \caption{Phase spaces as the center of mass is moved relative to the center of mobility along the tan sequence of points in Figure~\ref{fig:map}(a). (a-d) demonstrate a series of particles which pass through a Hopf bifurcation (a-b), create a stable and unstable limit cycle (b-c), which then disappear in a homoclinic bifurcation (c-d).}
    \label{fig:tan}
\end{figure*}

Particles with center of force positions given by the tan trajectory in Figure~\ref{fig:map}(a), yield phase spaces shown in Figure~\ref{fig:tan}.  We observe stable (blue) and unstable (red) limit cycles (b-c) flanked on either side by Hopf and homoclinic bifurcations. At the Hopf bifurcation, two spirals switch stability and nucleate two limit cycles. At the homoclinic bifurcation, an unstable manifold of each saddle no longer converges to the limit cycle and instead meets the stable manifold in a homoclinic orbit beyond which the limit cycle vanishes. 

We can define three-dimensional surfaces where each of these bifurcations occur within the alignment bifurcation shell. Figure~\ref{fig:map}(a) shows the region where limit cycles occur in grey, bounded by Hopf (blue) and homoclinic (red) bifurcation surfaces. The three-dimensional structure of the Hopf bifurcation surface is also shown in Figure~\ref{fig:surface}(b). More work is needed to determine how the size and shape of the regions of parameter space with limit cycles changes for particles with different shapes. 

Limit cycles are mentioned as a possibility by \citet{witten_review_2020}. For helical ribbons, limit cycles exist for a very narrow range of parameters and are difficult to realize experimentally.

\subsubsection{Fixed Point Swapping}
\begin{figure*}
    \centering
    \includegraphics[width=1\linewidth]{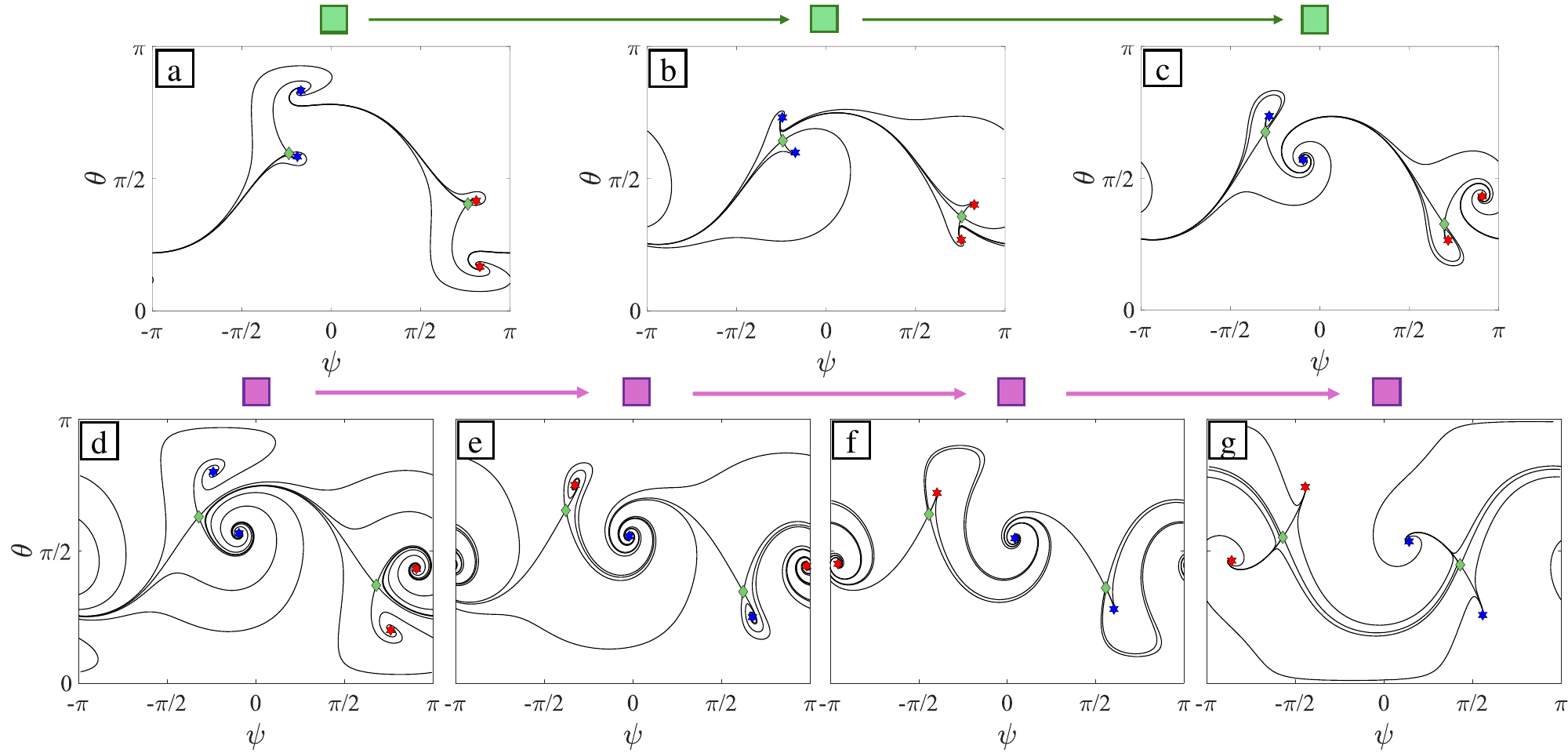}
    \caption{Phase spaces as the center of mass is moved relative to the center of mobility along the green and purple sequences of points in Figure~\ref{fig:map}(a). (a-c) demonstrate orientation dynamics above (a) within (b) and below (c) a cusp. (d-g) shows a longer trajectory in $\ve r$ where the attracting and repelling spirals swap stability (d-e) and swap partners (f-g).}
    \label{fig:greenpink}
\end{figure*}

As $\ve r$ increases and the particle approaches a 6 to 2 bifurcation, fixed points tend to pair off with the partner with which they will eventually annihilate. Offsets along each principal axis result in different pairs, so there must be some $\ve r$ where the annihilating pairs swap. We see this swap occur as we vary $\ve r$ along the green trajectory, phase spaces pictured in Figure \ref{fig:greenpink}(a-c), where we approach the cusp from either side. The spirals that annihilate with the saddle fixed points switch depending on which side of the cusp $\ve r$ lies on. Moving a particle's center of mass so that it passes directly through the cusp in the 6 to 2 bifurcation surface results in all 3 fixed points merging to become a single spiral.  

\citet{vaquero2025fluttering} observe this type of bifurcation as they introduce geometric asymmetry in bent disks by bending them around cones, which breaks the fore-aft symmetry associated with one of the reflection planes. In $\ma b_m$'s eigenframe this corresponds to adding a component of $\ve r$ along the angle bisector of $\ma b_m$'s major and minor axes, which seems to be the location of a cusp in their particle's bifurcation surface. Note, that this particle has the $b_m$ eigenframe rotated by $\pi/4$ from the $\ma c$ eigenframe, so the bifurcation surface is different from the helical ribbon.   Also, they are changing the particle geometry together with its center of mass offset.

Finally, we vary $\ve r$ from the top to the bottom of the 6 to 2 bifurcation surface, along the pink trajectory in Figure~\ref{fig:greenpink}(a), with phase spaces shown in Figure~\ref{fig:greenpink}(d-g). Figure~\ref{fig:greenpink}(d) is near one cusp where the left saddle is symmetric between two stable spirals.  Between (d) and (e) one of these switches to an unstable spiral as a consequence of a hopf and homoclinic bifurcation.  Figure~\ref{fig:greenpink}(f) continues near the 6 to 2 bifurcation and by Figure~\ref{fig:greenpink}(g) it is approaching the other cusp where two unstable spirals are symmetrically near the left saddle.     Overall, 2 spirals swap stability and the other 2 swap position.

\section{Conclusion}

In this work, we introduce a unifying perspective from which  to understand the possible dynamics of sedimenting particles with shapes that couple translation to rotation.  The key idea is that any particle shape has a reference particle of the same shape and with the center of force coincident with the center of mobility.  This cocentered particle could be made with a special density distribution.   A cocentered particle has simple dynamics with only closed orbits around 4 centers and 2 saddles.  We explain these dynamics as a consequence of three planes of PT symmetry that are perpendicular to the orthogonal eigenvectors of the $\ma b_m$ tensor.   The dynamics of more general particles can be understood as a result of moving the center of force away from the center of mobility.  

We performed experiments where we changed the density distribution of helical ribbons so that the center of force moved along each of the principal axes, but the shape of the particle did not change.  Offsets along a particle's principal axes, defined by the shared eigenvectors of $\ma b_m$ and $\ma c$, preserve one PT symmetry.   Along principal axes associated with the major and minor eigenvalues of $\ma b_m$, an infinitesimal offset turns two of the centers into spirals and some trajectories approach a fixed orientation in the 2D orientation phase space while others remain in closed orbits.  An important conclusion is that often sedimentation dynamics are not a property of the particle alone, but also depend on the initial orientation. For small offsets along the intermediate principal axis, all trajectories remain in closed orbits.  For larger offsets in all directions, there is  a double saddle-node bifurcation which leaves behind one stable and one unstable orientation.

Numerical simulations using the immersed boundary method provided mobility tensors that were used to simulate sedimentation dynamics of helical ribbons.  These simulations show the same bifurcations observed in the experiments and they occur at similar values of the center of mass offset.  These simulations allow more detailed study and find a fascinating line of nodes bifurcation that allows the bifurcations along the intermediate axis to conserve topological index.  

Offsets simulated in arbitrary directions result in the same eventual bifurcation from 6 to 2 fixed points. We visualize the 6 to 2 bifurcation as a surface in the 3D space defined by the offset vector $\ve r$ which we call the alignment bifurcation surface. Using numerical simulations, we vary $\ve r$ within the alignment bifurcation surface and discover limit cycles flanked by homoclinic and Hopf bifurcations. These two bifurcations define additional surfaces embedded within the larger alignment bifurcation surface. 

We find that particles are extremely sensitive to changes in their center of mass position. Center of mass offsets on the order of 3D printing tolerance can result in limit cycles, convergence, and orbits that coexist in the orientation phase space.   Offsets of less than 1 percent of the length of our helical ribbons move the center of force outside the alignment bifurcation surface.   The scaling of the size of the alignment bifurcation surface is set by the length scale $| \ma b_m| /|\ma c|$, and the reason for the small size of the alignment bifurcation surface can be understood as a consequence of the fact that this length scale is much smaller than a typical dimension of a particle.  Or in other words, the non-dimensional translation-rotation coupling is small compared with the non-dimensional rotational drag.
The extreme sensitivity of a particle's orientation dynamics to density inhomogeneity is a useful feature for biological, or artificial, swimmers interested in controlling their orientation relative to the fluid environment with minimal actuation. 

This new perspective, which frames particle sedimentation dynamics in the space of offsets between the center of mass and center of mobility, offers new approaches to the helical design problem.  Sometimes called the chiral design problem, this is the problem of choosing geometric shapes that have desired dynamics due to translation-rotation coupling.    From this perspective, a primary focus should be on the eigenvalues of the translation-rotation coupling tensor about the center of mobility.  This will choose the class of cocentered particle.  Then optimizing the eigenvalues will optimize the cocentered translation coupling which can be followed by moving the center of mass as desired.

A set of fundamental questions remain unanswered concerning shapes where the eigenvectors of $\ma b_m$ are not aligned with the eigenvectors of $\ma c$.
We have studied helical ribbons for which these two eigenframes are the same.  Particles such as bent disks~\citep{miara_dynamics_2024}, other di-bilaterals~\citep{joshi2025sedimentation}, or curved wires~\citep{candalier2025curvedfibres} have $\ma b_m$ and $\ma c$ which share one eigenvector but not the other two.  Our analylsis captures the dynamics observed for these particles because the reflection symmetries ensure that the center of mass offset occurs along the shared eigenvector.  However, other center of mass offsets for these particles as well as shapes with completely different eigenframes for $\ma b_m$ and $\ma c$ will have different dynamics.   The rotation between these two eigenframes is an additional target for helical design.

\section{Acknowledgments}

 We thank Samitha Sanjuka for assistance with the experiments.  We thank Federico Toschi and Xander de Wit for the fruitful discussions and help with the simulations to determine the mobility tensors. This work was partially supported by NSF grant DMR-1508575, Army Research Office grant W911NF-17-1-0176,
Vetenskapsr\aa{}det grant no. 2021-4452 and the project “Shaping turbulence with smart particles” with Project No. OCENW.GROOT.2019.031 of the research program Open Competitie ENW XL, which is (partly) financed by the Dutch Research Council (NWO). Computational resources were provided by the Netherlands Organization for Scientific Research (NWO) through the use of supercomputer facilities (Snellius) under Grant No. 2023.026.  BM acknowledges support from an EAISI visiting professorship at the TU Eindhoven.

\section{Declaration of Interests} The authors report no conflict of interest.

\section{Appendix}

\subsection{Fixed Point Stability for Cocentered Particles}
\label{fpstability1}
For the cocentered case where $\ma b_f=\ma b_m$ is symmetric, it is straight forward to evaluate  the Jacobian of the dynamical system defined by      (\ref{eq:g}) in the eigenframe of $\ma b_m$. 
\begin{equation}
    \ma J = \nabla_g \frac{d \hat{\ve g}}{dt}=
    \begin{bmatrix}
        0 & (b_3 - b_2) g_3 & (b_3 - b_2) g_2\\
        (b_1 - b_3) g_3 & 0 & (b_1 - b_3) g_1 \\
        (b_2 - b_1) g_2 & (b_2 - b_1) g_1& 0
    \end{bmatrix}
    \label{Jacobian}
\end{equation}
where $b_1\leq b_2\leq b_3$ are the eigenvalues of $\ma b_m$ ordered from smallest to largest and $g_i$ are the components of the gravity vector in the body frame that is the eigenframe of $\ma b_m$.  Fixed points occur where gravity is aligned with an eigenvector of $\ma b_m$. The magnitude of $\hat{\ve g}$ is fixed, so the dynamics are two dimensional on a sphere and we can evaluate $\ma J$ in the 2D tangent space at each fixed point, which we call $\ma J^*$. Each fixed point has $\tr (\ma J^*) = 0$, so $\det(\ma J^*)$ determines if each fixed point is a saddle or a center.

When \ma J is evaluated at the fixed point along the minor axis with only $g_1$ nonzero, we have 
\begin{eqnarray}
    \det(\ma J^*) = (b_3 -b_1)(b_2 -b_1)\geq 0
\end{eqnarray}

When \ma J is evaluated at the fixed point along the major axis with only $g_3$ nonzero, we have
\begin{eqnarray}
    \det(\ma J^*) = (b_3 -b_2)(b_3 -b_1)\geq 0
\end{eqnarray}
So, the 4 fixed points along the minor and major axes of $\ma b_f$ are centers.

When $\ma J^*$ is evaluated at the fixed point along the intermediate axis with only $g_2$ nonzero, we have
\begin{eqnarray}
    \det(\ma J^*) = (b _3 -b_2)(b_1 -b_2)\leq 0
\end{eqnarray}
So, the 2 fixed points along the intermediate axis of $\ma b_m$ are saddles. 
This property that the intermediate axis of a triaxial cocentered particle which couples translation to rotation is quite similar to the intermediate axis theorem in classical mechanics, which states that for solid body rotations, the fixed point along the intermediate axis of the of rotational inertia tensor is a saddle. 

\subsection{Fixed Point Stability for Finite Center of Mass Offsets}
\label{fpstability2}

For non-zero $\ve r$, the Jacobian is

\begin{equation}
    \ma J = \nabla_g \frac{d \hat{\ve g}}{dt} = \nabla_g(-(\ma b_m + \ma c \times \ve  r) \hat{\ve g}) \times \hat{\ve g}
\end{equation}

which can be computed to be

\begin{equation}
\begin{aligned}
\ma J
&=
    \begin{bmatrix}
        0 & (b_3 - b_2) g_3 & (b_3 - b_2) g_2\\
        (b_1 - b_3) g_3 & 0 & (b_1 - b_3) g_1 \\
        (b_2 - b_1) g_2 & (b_2 - b_1) g_1& 0
        \end{bmatrix}
\\[6pt]
&\quad+
\begin{bmatrix}
- c_3 g_2 r_2 - c_2 g_3 r_3
&  c_3 (2g_2r_1 - g_1 r_2) 
&  c_2( 2 g_3 r_1 - g_1 r_3 )
\\
c_3(2 g_1 r_2 - g_2 r_1)
& - c_3g_1 r_1 -c_1 g_3 r_3
&  c_1(2 g_3 r_2 -  g_2 r_3
\\
c_2(2 g_1 r_3 - g_3 r_1)
& c_1(2 g_2 r_3 - g_3 r_2)
& - c_2 g_1 r_1 - c_1 g_2 r_2
\end{bmatrix}.
\end{aligned}
\label{jacobian_r}
\end{equation}\\
We consider the case where $\ve r$ lies along an eigenvector of $\ma b_m$, and examine the fixed point along $\ve r$, for which (\ref{jacobian_r}) simplifies considerably.  The other fixed points move as $\ve r$ changes so they are not simple to study analytically.
When $\ve r$ lies along the minor axis of $\ma b_m$, with $\ve r = (r_1,0,0)$, we can evaluate $\ma J$ at $\hat{\ve g} = (1,0,0)$
\begin{equation}
    \ma J = \begin{bmatrix}
0 & 0 &0\\
0 & -c_3r_1 & b_1 - b_3  \\
0 & b_2 - b_1 & -c_2r_1
\end{bmatrix}
\end{equation}\\
and can calculate the determinant and trace in the 2D tangent space to be

\begin{equation}
\begin{aligned}
\det \ma J^* = (r_1)^2c_2c_3 -(b_1 -b_3)(b_2 -b_1) \qquad\qquad
\tr \ma J^* = -r_1(c_2 + c_3)
\end{aligned}
\end{equation}\\
When $\ve r$ lies along the major axis of $\ma b_m$, with $\ve r = (0,0,r_3)$, we can evaluate $\ma J$ at $\hat{\ve g} = (0,0,1)$

\begin{equation}
    \ma J = \begin{bmatrix}
-c_2r_3 & b_3 - b_2  & 0\\
b_1 - b_3  & -c_1r_3 & 0 \\
0 & 0& 0
\end{bmatrix}
\end{equation}\\
and can calculate the 2D determinant and trace to be
\begin{equation}
\begin{aligned}
\det \ma J^* = (r_3)^2c_1c_2 - (b_3 -b_2)(b_1 -b_3) \qquad\qquad
\tr \ma J^* = - r_3(c_1 + c_2)
\end{aligned}
\end{equation}\\
When $\ve r$ lies along the intermediate axis of $\ma b_m$, with $\ve r = (0,r_2,0)$, we can evaluate $\ma J$ at $\hat{\ve g} = (0,1,0)$
\begin{equation}
    \ma J = \begin{bmatrix}
-c_3r_2 & 0 & b_3 - b_2 \\
0 & 0 & 0\\
b_2 - b_1  & 0& -c_1r_2
\end{bmatrix}
\end{equation}\\
and can calculate the 2D determinant and trace to be

\begin{equation}
\begin{aligned}
\det \ma J^* = (r_2)^2c_1c_3 - (b_3 -b_2)(b_2 -b_1) \qquad\qquad
\tr \ma J^* = -r_2(c_1+c_3)
\end{aligned}
\end{equation}

In each case, the introduction of a center of mass offset adds a positive term to $\det \ma J^*$, because rotational mobility must have a positive value and so $\ma c$ is positive definite. Therefore, for offsets along the major and minor axes $\det \ma J^* > 0$ for all $|\ve r|$, and the fixed points cannot be saddles. The sign of $\tr \ma J^*$ depends on the relative sign of $\hat{\ve g}$ and $\ve r$, but is always nonzero, which excludes centers. Intuitively, the fixed point aligned with $\ve r$ is stable ($\tr \ma J^* < 0$), and the anti-aligned fixed point is unstable ($\tr \ma J^* > 0$).

Offsets along the intermediate axis of $\ma b_m$ are slightly different. When $\ve r = 0$, $\det \ma J^* < 0$ and the corresponding fixed points are saddles (derived in Appendix 1). As $|r_2|$ increases, $\det \ma J^*$ becomes less negative until it swaps signs at a special offset magnitude $r^*_2$, at which point the fixed point becomes stable ($\tr \ma J^* < 0$) or unstable ($\tr \ma J^* > 0$). The critical value $r^*_2$ can be calculated by setting $\det \ma J = 0$ which gives

\begin{eqnarray}
    r^*_2 = \pm\sqrt{\frac{(b_3-b_2)(b_2-b_1)}{c_1c_3}}
    \label{eq:surfscale}
\end{eqnarray}

We discuss the implications of this expression for $r_2^*$ in the main text.

\subsection{Translation Theorem Under Orthogonal Similarity Transformations}
\label{ortho_translationtheorem_appendix}

We have the identity

\begin{equation}
     \tau \ma b_f =\det(\ma Q)\ma Q \ma b_f \ma Q^{-1} 
\label{sym_appendix}
\end{equation}

We can use the translation theorem

\begin{equation}
\ma b_{f} = \ma b_{m}+ \ma c \times\ve r
\label{trans_appendix}
\end{equation}

to make (\ref{sym_appendix}) a condition on $\ve r$, $\ma c$, and $\ma b_m$. Plugging (\ref{trans_appendix}) into (\ref{sym_appendix}), we have

\begin{equation}
    \tau\ma b_{m}+ \tau(\ma c \times\ve r) =\det(\ma Q)\ma Q\ma b_{m}\ma Q^{-1}+\det(\ma Q)\ma Q\ma (\ma c \times\ve r)\ma Q^{-1} 
    \label{full_appendix}
\end{equation}

We adopt the notation $[\ve r]_{ij} =\epsilon_{ijk}r_k$, where $\epsilon_{ijk}$ is the Levi-Civita tensor, to denote the antisymmetric matrix built from a vector $\ve r$

\begin{equation}
    [\ve r] = \begin{bmatrix}
        0 &-r_3 & r_2\\
        r_3 & 0 & -r_1\\
        -r_2 & r_1& 0
    \end{bmatrix}
\end{equation}

defined so that $\ma c \times \ve r = -[\ve r] \ma c$ for some vector $\ve r$ and tensor $\ma c$. In this notation, it is easy to check that the columns of $\ma c \times \ve r$ are the cross products between each column of $\ma c$ and the vector $\ve r$.

We can rewrite the second term on the right side of (\ref{full_appendix}) as

\begin{equation}
\begin{aligned}
\det(\ma Q)\ma Q\ma (\ma c \times\ve r)\ma Q^{-1} &=-\det(\ma Q)\ma Q\ma ([\ve r]\ma c)\ma Q^{-1}  \\
&= -\det(\ma Q)(\ma Q\ma [\ve r]\ma Q^{-1})(\ma Q \ma c\ma Q^{-1})
\end{aligned}
    \label{identity1}
\end{equation}

Using index notation, and the definition of $[\ve r]$, we have

\begin{equation}
\begin{aligned}
(\ma Q[\ve r] \ma Q^{-1})_{ij}
&= Q_{ia} Q_{jb}\,\epsilon_{abc}\,r_c \\
&= Q_{ia} Q_{jb}\,(Q_{kc}Q_{k\ell})\,\epsilon_{ab\ell}\,r_c \\
&= Q_{ia} Q_{jb} Q_{k\ell}\,\epsilon_{ab\ell}\,(Q_{kc} r_c) \\
&= \det(Q)\,\epsilon_{ijk}\,(Q_{kc} r_c) \\
&= \det(\ma Q)\,[\ma Q \ve r]_{ij}.
\end{aligned}
\label{identity_appendix}
\end{equation}

Where we've used the definition of an orthogonal matrix $Q_{kc}Q_{kl}=\delta_{cl}$ and the Levi-Civita identity for non-singular $\ma Q$

\begin{equation}
Q_{ia}Q_{jb}Q_{kl}\epsilon_{abl}=\det(Q)\epsilon_{ijk}
\end{equation}

Inserting (\ref{identity_appendix}) into (\ref{identity1}) we have

\begin{equation}
\begin{aligned}
\det(\ma Q)\ma Q\ma (\ma c \times\ve r)\ma Q^{-1} &=-\det(\ma Q)^2 [\ma Q \ve r](\ma Q \ma c\ma Q^{-1})  \\
&= (\ma Q \ma c\ma Q^{-1}) \times (\ma Q \ve r)
\end{aligned}
\end{equation}

which can be substituted into (\ref{full_appendix}) to get

\begin{equation}
    \tau\ma b_{m}+ \tau(\ma c \times\ve r) =\det(\ma Q)\ma Q\ma b_{m}\ma Q^{-1}+(\ma Q \ma c\ma Q^{-1}) \times (\ma Q \ve r)
\end{equation}

\subsection{Computing Center of Mass of 3D Printed Helical Ribbons}
\label{Exp_appendix}

The center of mass of a system of discrete objects is
\begin{equation}
\label{eq_rcomtotal_1}
    \ve r_{cm}=\frac{\Sigma \ve r_i m_i}{\Sigma m_i}. 
\end{equation}
where $m_i$ is the mass and $\ve r_i$ is the center of mass of object $i$.   We include four objects. (1) a solid 3D printed plastic helical ribbon with no hole that has mass  $m_p$ and center of mass $\ve r_p$, (2) the hole in plastic with mass $m_h$ which was removed to create the hole for the bead to sit and is at position $\ve r_h$, 
(3) the metal sphere with mass $m_b$ at the position of the hole $\ve r_h$.   (4) the epoxy with mass $m_e$ whose center of mass is also at the hole, $\ve r_h$.  
The center of mass of the composite object is then 
\begin{equation}
\label{eq_rcomtotal_2}
    \ve r_{cm}=\frac{\ve r_p m_p+\ve r_h(m_b+m_e-m_h)}{m_p+m_b+m_e-m_h},
\end{equation}

However, we do not have an accurate way to measure $\ve r_p$.  It would be zero, except that the 3D printer produces particles with a small unknown center of mass offset.   We use what we know about the cocentered case to fit to determine $\ve r_p$. For a cocentered particle, $\ve r_{cm}=0$, and therefore we can re-write (\ref{eq_rcomtotal_2}) as  
\begin{equation}
\label{eq_rcom_cocenter}
    0=\frac{\ve r_p m_p+\ve r_{cc}(m_s+m_e-m_h)}{m_p+m_s+m_e-m_h},
\end{equation}
where $\ve r_{cc}$ is the location where a steel bead with mass $m_s$ must be placed in order to counteract the density inhomogeneity due to our 3D printer. To find $\ve r_{cc}$, we used particles with steel spheres added at different known locations that spanned the cocentered case along each of the three principal axes.  Measurements of the asymmetry of the phase space dynamics were interpolated to find $\ve r_{cc}$ where the dynamics were cocentered.   With measured $\ve r_{cc}$, we can obtain the position of the center of mass of the 3D printed plastic ribbon,
\begin{equation}
\label{eq_rcom_solid}
    \ve r_p=\frac{-\ve r_{cc}(m_s+m_e-m_h)}{m_p}.
\end{equation}
From (\ref{eq_rcomtotal_2}), we then obtain an expression for the center of mass of the composite particles used in the experiment using only position vectors and masses that we can measure:
\begin{equation}
\label{eq_rcom_final}
    \ve r_{cm}=\frac{-\ve r_{cc}(m_s+m_e-m_h)+\ve r_b(m_b+m_e-m_h)}{m_p+m_b+m_e-m_h}.
\end{equation}



\subsection{Supplementary videos}
\label{supplementary_videos}
Animations of phase space dynamics as the center of mass is moved along the principal axes of $\ma b_m$ are available at:  \url{https://youtu.be/VZcN-ursxY4} for center of mass offsets in the x direction which has the intermediate eigenvalue of $\ma b_m$, \url{https://youtu.be/6y1DfvOqERo} for center of mass offsets in the y direction which has the minor eigenvalue  of $\ma b_m$, and \url{https://youtu.be/ll2p49CXIeI} for center of mass offsets in the z direction which has the major eigenvalue  of $\ma b_m$. A collection of all animations is accessible at: \url{https://www.youtube.com/playlist?list=PLr9-yt1AsSrIjZPdM3YQTqsqBHrqq-C_9}.

\bibliographystyle{jfm}
\bibliography{bibFINAL}

\end{document}